\PassOptionsToPackage{pdfpagelabels=false}{hyperref}
\documentclass[useAMS,usenatbib]{mn2e}
\def \eg {e.g.}

\def \ie {i.e.}
\def \cf {cf.}
\def \lcdm {{\hbox{$\Lambda$CDM}}}
\def \omegam {{\hbox{$\Omega_m$}}}
\def \omegal {{\hbox{$\Omega_\Lambda$}}}
\def \hzero {{\hbox{$H_0$}}}
\def \arcmin {\hbox{$^\prime$}}
\def \arcsec {\hbox{$^{\prime\prime}$}}
\def \deg {\hbox{$^\circ$}}
\def \nh {\hbox{$N_{\rm H}$}}

\def \mach {{\hbox{$\mathcal{M}$}}}
\def \machkt {{\hbox{$\mathcal{M}_{\rm kT}$}}}

\def \cstatdof {\hbox{c-stat/d.o.f.}}
\def \msun {\hbox{${\rm M_\odot}$}}
\def \zsun {\hbox{${\rm Z_\odot}$}}

\def \mfive {\hbox{$M_{500}$}}

\def \rfive {\hbox{$r_{500}$}}

\newcommand{\kmsmpc }{\mbox{km s$^{-1}$ Mpc$^{-1}$}}

\newcommand{\kev }{\mbox{keV}}

\newcommand{\jy }{\mbox{Jy}}
\newcommand{\mujyb }{\mbox{$\mu$Jy beam$^{-1}$}}
\newcommand{\muG }{\mbox{$\mu$G}}
\newcommand{\whz }{\mbox{W Hz$^{-1}$}}
\newcommand{\acisi }{ACIS-I}
\newcommand{\aciss }{ACIS-S}

\newcommand{\obsid }{ObsID}

\newcommand{\uv }{\textit{uv}}
\newcommand{\aips }{\textsc{aips}}
\newcommand{\aipsE }{Astronomical Image Processing System}
\newcommand{\factor }{\textsc{factor}}
\newcommand{\prefactor }{\textsc{prefactor}}
\newcommand{\awimager }{\textsc{AWimager}}
\newcommand{\wsclean }{\textsc{WSClean}}
\newcommand{\casa }{\textsc{casa}}

\newcommand{\spam }{\textsc{spam}}
\newcommand{\spamE }{Source Peeling and Atmospheric Modeling}
\newcommand{\pybdsf }{\textsc{pybdsf}}
\newcommand{\pybdsfE }{PYthon Blob Detector and Source Finder}
\newcommand{\xspec }{\textsc{xspec}}
\newcommand{\contbin }{\textsc{contbin}}
\newcommand{\ciao }{\textsc{ciao}}

\newcommand{\caldb }{\textsc{caldb}}

\newcommand{\xmm }{{\em XMM-Newton}}
\newcommand{\chandra }{{\em Chandra}}

\newcommand{\planck }{{\em Planck}}

\newcommand{\rosat }{ROSAT}

\newcommand{\gmrt }{GMRT}
\newcommand{\gmrtE }{Giant Metrewave Radio Telescope}

\newcommand{\vla }{VLA}
\newcommand{\vlaE }{Very Large Array}

\newcommand{\lofar }{LOFAR}
\newcommand{\lofarE }{LOw Frequency ARray}

\newcommand{\wsrt }{WSRT}
\newcommand{\wsrtE }{Westerbork Synthesis Radio Telescope}

\newcommand{\lotss }{LoTSS}
\newcommand{\lotssE }{LOFAR Two-meter Sky Survey}

\newcommand{\tgss }{TGSS}
\newcommand{\tgssE }{TIFR GMRT Sky Survey}

\newcommand{\nvss }{NVSS}
\newcommand{\nvssE }{NRAO VLA Sky Survey}
\newcommand{\wenss }{WENSS}
\newcommand{\wenssE }{WEsterbork Northern Sky Survey}
\newcommand{\vlss }{VLSS}
\newcommand{\vlssE }{VLA Low-frequency Sky Survey}

\newcommand{\sdss }{SDSS}

\usepackage{xcolor}
\usepackage{graphicx}
\usepackage{amssymb}
\usepackage{multirow,bigdelim}
\usepackage[export]{adjustbox}
\defcitealias{giovannini09}{G09}
\defcitealias{venturi13}{V13}
\title[Discovery of a double radio halo in Abell 1758]{LOFAR discovery of a double radio halo system in Abell 1758 and radio/X-ray study of the cluster pair}
\author[Botteon et al.]{%
A.~Botteon$^{1,2}$\thanks{E-mail: botteon@ira.inaf.it}, T.~W.~Shimwell$^{3,4}$, A.~Bonafede$^{1,2,5}$, D.~Dallacasa$^{1,2}$, G.~Brunetti$^{2}$, \newauthor
S.~Mandal$^{4}$, R.~J.~van Weeren$^{4}$, M.~Br\"{u}ggen$^{5}$, R.~Cassano$^{2}$, F.~de Gasperin$^{4}$, \newauthor
D.~N.~Hoang$^{4}$, M.~Hoeft$^{6}$, H.~J.~A.~R\"{o}ttgering$^{4}$, F.~Savini$^{5}$, G.~J.~White$^{7,8}$,
\newauthor A.~Wilber$^{5}$ and T.~Venturi$^{2}$ \\
$^{1}$Dipartimento di Fisica e Astronomia, Universit\`{a} di Bologna, via P.~Gobetti 93/2, I-40129 Bologna, Italy \\
$^{2}$INAF - IRA, via P.~Gobetti 101, I-40129 Bologna, Italy \\
$^{3}$ASTRON, the Netherlands Institute for Radio Astronomy, Postbus 2, NL-7990 AA Dwingeloo, The Netherlands \\
$^{4}$Leiden Observatory, Leiden University, PO Box 9513, NL-2300 RA Leiden, The Netherlands  \\
$^{5}$Hamburger Sternwarte, Universit\"{a}t Hamburg, Gojenbergsweg 112, D-21029 Hamburg, Germany \\
$^{6}$Th\"{u}ringer Landessternwarte, Sternwarte 5, D-07778 Tautenburg, Germany \\
$^{7}$Department of Physical Sciences, The Open University, Milton Keynes MK7 6AA, England \\
$^{8}$Space Science and Technology Department, The Rutherford Appleton Laboratory, Chilton, Didcot, Oxfordshire OX11 0NL 
}
\date{\today}

\date{Accepted XXX. Received YYY; in original form ZZZ}

\pubyear{2018}

\begin{document}
\label{firstpage}
\pagerange{\pageref{firstpage}--\pageref{lastpage}}
\maketitle

\begin{abstract}
Radio halos and radio relics are diffuse synchrotron sources that extend over Mpc-scales and are found in a number of merger galaxy clusters. They are believed to form as a consequence of the energy that is dissipated by turbulence and shocks in the intra-cluster medium (ICM). However, the precise physical processes that generate these steep synchrotron spectrum sources are still poorly constrained. We present a new \lofar\ observation of the double galaxy cluster Abell 1758. This system is composed of A1758N, a massive cluster hosting a known giant radio halo, and A1758S, which is a less massive cluster whose diffuse radio emission is confirmed here for the first time. Our observations have revealed a radio halo and a candidate radio relic in A1758S, and a suggestion of emission along the bridge connecting the two systems which deserves confirmation. We combined the \lofar\ data with archival \vla\ and \gmrt\ observations to constrain the spectral properties of the diffuse emission. We also analyzed a deep archival \chandra\ observation and used this to provide evidence that A1758N and A1758S are in a pre-merger phase. The ICM temperature across the bridge that connects the two systems shows a jump which might indicate the presence of a transversal shock generated in the initial stage of the merger.
\end{abstract}


\section{Introduction}

Diffuse and low surface brightness radio emission with steep spectrum ($\alpha>1$, with $S_\nu \propto \nu^{-\alpha}$) associated with the intra-cluster medium (ICM) has been found in a number of merging galaxy clusters \citep[see][for an overview]{feretti12rev}. This emission, generally referred to as radio halos or radio relics, probes the synchrotron radiation from $\sim$ GeV relativistic electrons in the $\sim$ \muG\ cluster-scale magnetic fields. \\
\indent
Over the last two decades, important steps forward have been made in understanding the origin of non-thermal phenomena in the ICM. A milestone in this framework was the connection between diffuse radio emission in the ICM and dynamically disturbed systems \citep[\eg][]{buote01, cassano10connection, cassano13, cuciti15}. This suggests that cluster mergers play a crucial role in the formation of halos and relics. During mergers, turbulence and shocks are generated in the ICM where they dissipate energy. A small fraction of this energy goes into the re-acceleration of relativistic particles and the amplification of the magnetic field \citep[see][for a review]{brunetti14rev}. \\
\indent
In particular, giant radio halos are thought to be generated by (primary or secondary) electrons re-accelerated by turbulent-driven mechanisms during mergers \citep[\eg][]{petrosian01, brunetti01coma, brunetti07cr, brunetti11, brunetti16stochastic, pinzke17, brunetti17}, although many aspects of the physics of these mechanisms remain poorly understood \citep[see][]{brunetti16challenge}. Radio relics are thought to be generated by electrons accelerated or re-accelerated by shocks \citep[\eg][]{ensslin98, roettiger99a3667, hoeft07, kang11, kang16reacc, kang12, pinzke13}. Whilst a connection between radio relics and merger shocks is fairly well established \citep[\eg][]{macario11, akamatsu13systematic, botteon16gordo}, open problems include the efficiency of particle acceleration and the proton-to-electron ratio in the acceleration \citep[\eg][]{vazza14challenge, vazza15efficiency, vazza16, kang14, guo14a, guo14b, wittor17relics}. Shock re-acceleration for radio relics is an emerging scenario that seems to be supported by recent observations of clusters where radio galaxies located close to the relic position can provide seed electrons that are more easily re-accelerated by low Mach number shocks. This alleviates the large requirements of acceleration efficiencies \citep{bonafede14reacc, shimwell15, botteon16a115, vanweeren17a3411}. \\
\indent
The \lofarE\ \citep[\lofar;][]{vanhaarlem13} is a radio interferometer observing in the range between 10 and 90~MHz with Low Band Antennas (LBA) and between 110 and 240~MHz with High Band Antennas (HBA). It has long been recognized that \lofar\ has the potential to make breakthroughs in the field of galaxy cluster science \citep[\eg][]{rottgering06, rottgering11, cassano10lofar, nuza12}. The high resolution allows for the structures of halos and relics to be precisely characterized and isolated from the often severe contamination from other radio emitting sources in the vicinity. The dense core provides excellent surface brightness sensitivity, that coupled with the low observing frequencies allows for sensitive observations. \\
\indent
The \lotssE\ \citep[\lotss;][]{shimwell17} is designed to fully exploit the potential of \lofar\ and produce  high resolution ($\sim 5\arcsec$) and high sensitivity ($\sim 100$ \mujyb) images of the entire northern sky at $120-168$~MHz. As part of this survey we have observed the double galaxy cluster Abell 1758. In this paper, we present this new \lofar\ observation together with archival \gmrtE\ (\gmrt), \vlaE\ (\vla) and \chandra\ data. Using these data we have discovered and characterized the second double radio halo system known to date and we argue that the two clusters that constitute Abell 1758 are in a pre-merger state. \\
\indent
Throughout this paper, we assume a \lcdm\ cosmology with $\omegal = 0.7$, $\omegam = 0.3$ and $\hzero = 70$ \kmsmpc. At the cluster redshift ($z=0.279$) this corresponds to a luminosity distance of $D_L = 1428$ Mpc and to an angular to linear conversion scale of $4.233$ \arcsec/kpc. Quoted errors are at $1\sigma$ confidence level for one parameter, unless otherwise stated.

\section{The double cluster Abell 1758}\label{sec:a1758}

Abell 1758 (hereafter A1758) is a galaxy cluster located at $z=0.279$ that has been intensively studied in the literature. Early \rosat\ data \citep{rizza98} revealed that it consists of two components, A1758N (in the north) and A1758S (in the south), separated by a projected distance of $\sim8$ arcmin (about 2 Mpc). \citet{david04} estimated  virial radii of $2.6$ Mpc (for A1758N) and $2.2$ Mpc (for A1758S), indicating that each cluster is affected by the potential well of the other and that they are gravitationally bound. Despite this, no signs of significant interaction between A1758N and A1758S were found by \chandra\ and \xmm\ observations \citep{david04}. However, from X-ray and optical studies, it is clear that the two sub-clusters are undergoing their own distinct mergers, with A1758N in a late and A1758S in an early merger state \citep[\eg][]{david04, boschin12a1758}. This might also be reflected in the infrared luminosity of the galaxies of A1758N, which is almost two times larger than that of A1758S, suggesting different dynamical histories for the two clusters \citep{haines09}. Weak lensing studies indicate that A1758N has a bimodal mass distribution, while A1758S represents a single mass clump \citep{dahle02, okabe08, ragozzine12, monteirooliveira17a1758}. Individual mergers are possibly occurring near the plane of the sky for A1758N and close to the line of sight for A1758S. \\
\indent
So far, most studies have focused on A1758N which is more massive and hotter than A1758S \citep[\eg][]{david04}. The mass of A1758N has been estimated using several methods (\eg\ X-ray scaling relations, \citealt{david04}; weak lensing, \citealt{okabe08}; member galaxy dynamics, \citealt{boschin12a1758}; hydrostatic equilibrium, \citealt{martino14}), providing a virial mass of $\sim10^{15}$ \msun\ which is split approximately equally between the two sub-components. This is further supported by hydrodynamical simulations, which can reproduce the X-ray morphology of A1758N assuming an off-axis collision of two equal mass ($\sim 5 \times 10^{14}$ \msun) clusters \citep{machado15a1758}. A compilation of the different mass estimates reported for A1758N is given in Tab.~1 of \citet{monteirooliveira17a1758}. Note that the mass of A1758S is more uncertain whilst it appears to be at least a factor of 1.5 smaller than that of A1758N \citep{david04, ragozzine12, haines17arx}.  \\
\indent
In the radio band, A1758N hosts a giant radio halo that was first detected by \citet{kempner01} and later investigated at 1.4~GHz with the \vla\ \citep[][hereafter \citetalias{giovannini09}]{giovannini09} and at 325~MHz with the \gmrt\ \citep[][hereafter \citetalias{venturi13}]{venturi13}. There are no reports of diffuse radio emission associated with A1758S in the literature. 

\section{Observations and data reduction}

\begin{table*}
 \centering
 \caption{Summary of the radio observations used in this work.}
 \label{tab:radio_obs}
  \begin{tabular}{lcccc} 
  \hline
   & \lofar\ & \gmrt\ & \multicolumn{2}{c}{\vla} \\
   & & & Array C & Array D \\
  \hline
  Project code & LC2\_038 & 11TVA01 & \multicolumn{2}{c}{AG639} \\
  \multirow{2}*{Pointing center (RA, DEC)} & 13h37m30s & 13h32m32s & \multicolumn{2}{c}{13h32m32s} \\
   & +49\deg44\arcmin53\arcsec & +50\deg30\arcmin37\arcsec & \multicolumn{2}{c}{+50\deg30\arcmin36\arcsec}  \\
  Observation date & 2014 Jun 1 & 2007 Mar 30/31 & 2004 May 6 & 2003 Mar 11 \\
  Total on-source time (hr) & 8.0 & 8.0 & 2.5 & 2.5 \\
  Flux calibrator & 3C196 & 3C147 & \multicolumn{2}{c}{3C286} \\
  Total on-calibrator time (min) & 10 & 26 & 10 & 9 \\
  Central frequency (MHz) & 144 & 325 & \multicolumn{2}{c}{1425} \\
  Bandwidth (MHz) & 48 & 33 & \multicolumn{2}{c}{50} \\
  \hline
  \end{tabular}
\end{table*}

\begin{table*}
 \centering
 \caption{Imaging parameters for the radio images shown in the paper. The beam position angle (PA) is measured from north to east.}
 \label{tab:radio_imaging}
  \begin{tabular}{lrrrrrrr} 
  \hline
  Fig. & Instrument & Frequency & Robust & Taper & Resolution & PA & rms \\
   & & (MHz) & & (\arcsec) & ($\arcsec \times \arcsec$) & (\deg) & (\mujyb) \\
  \hline
  1 & \lofar\ & 144 & $+0.5$ & 0 & $34 \times 23$ & 45 & 230 \\
  1 & \gmrt\ & 325 & $0.0$ & 35 & $43 \times 29$ & 29 & 400 \\
  1 & \vla\ & 1425 & $0.0$ & 35 & $43 \times 38$ & 50 & 70 \\
  3 & \lofar\ & 144 & $0.0$ & 40 & $60 \times 51$ & 69 & 390 \\
  4 & \lofar\ & 144 & $-0.5$ & 10 & $16 \times 11$ & 88 & 140 \\
  \hline
  \end{tabular}
\end{table*}

\begin{figure*}
 \centering
 \includegraphics[width=.33\textwidth]{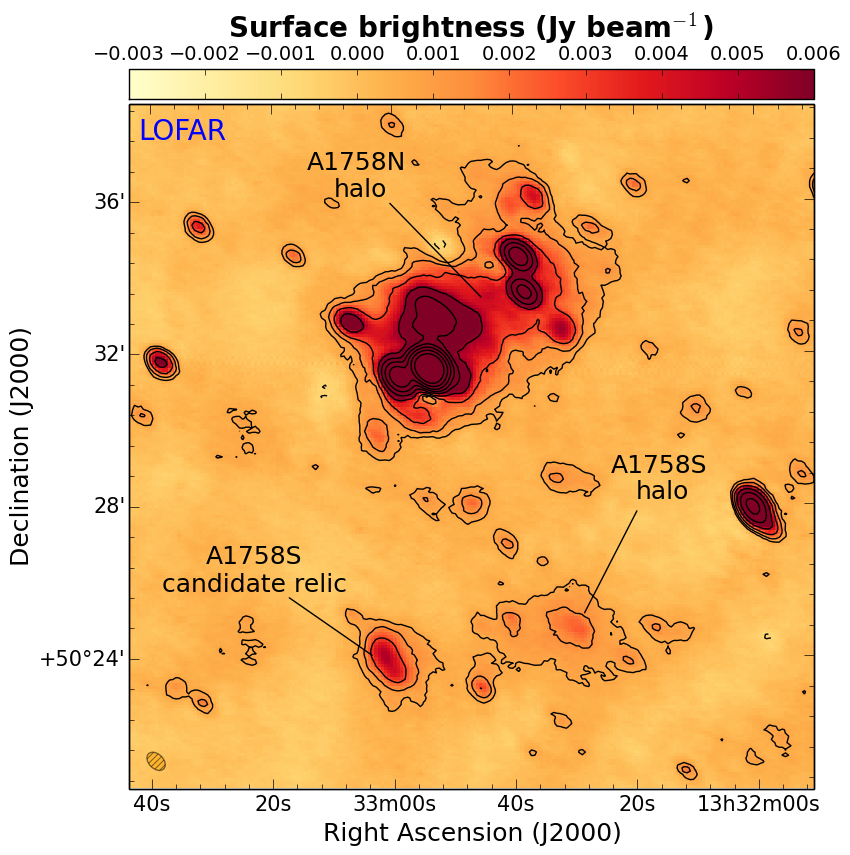}
 \includegraphics[width=.33\textwidth]{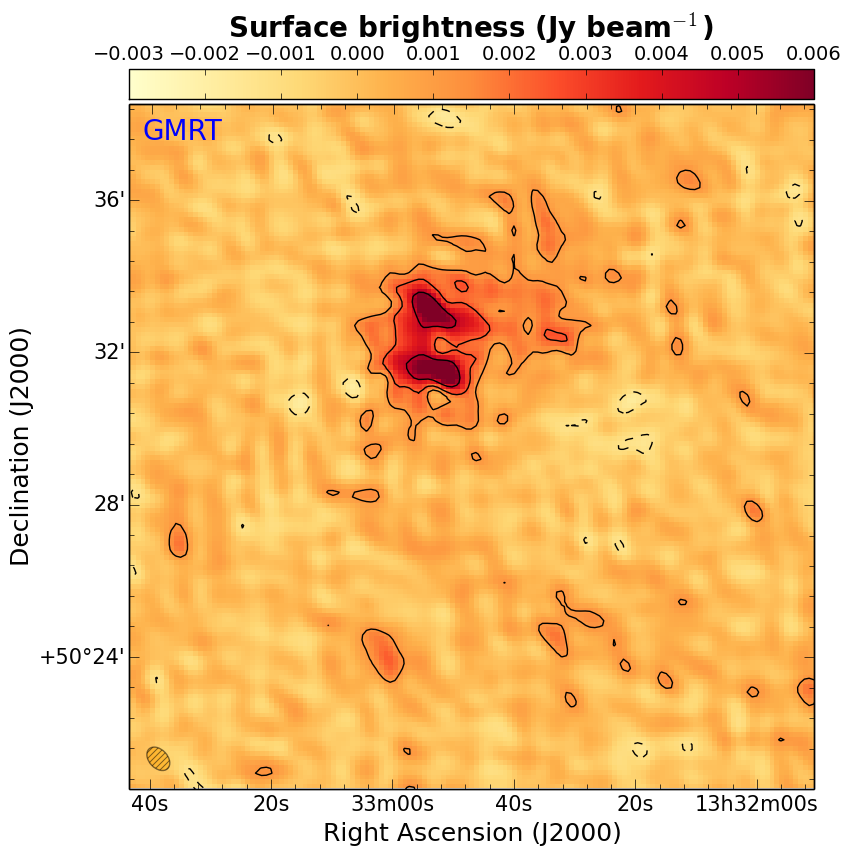}
 \includegraphics[width=.33\textwidth]{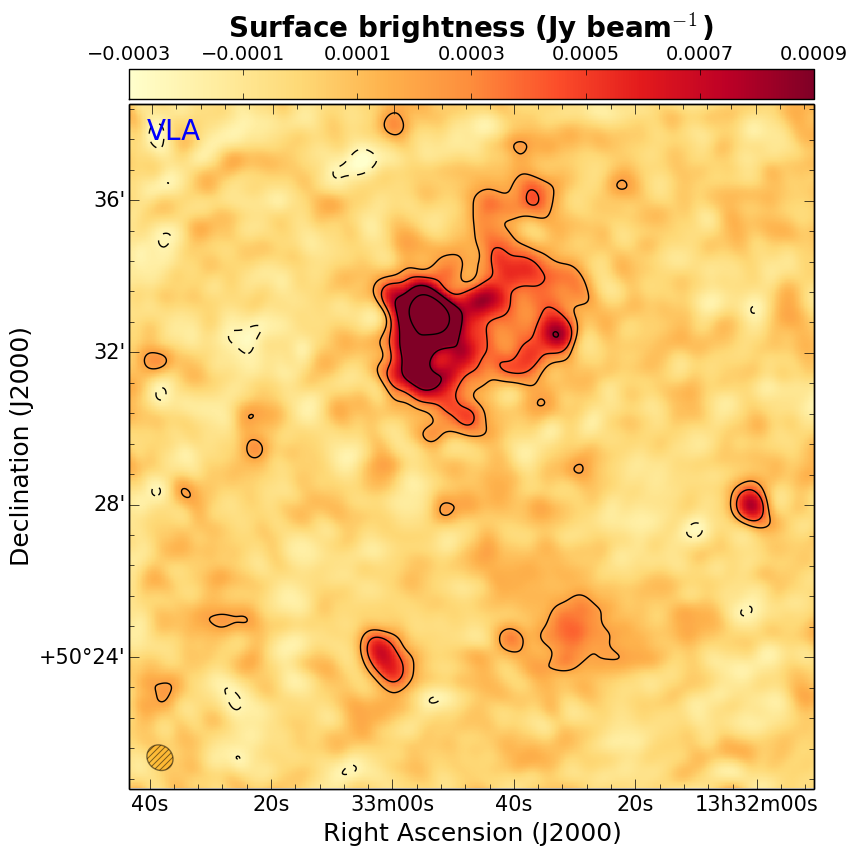}
 \caption{The cluster A1758 as observed in the radio band with \lofar\ (\textit{left}), \gmrt\ (\textit{center}) and \vla\ (\textit{right}). The point-sources were subtracted in the \gmrt\ and \vla\ images. Contours are spaced by a factor of 2 starting from $3\sigma$, where $\sigma_{\rm LOFAR} = 230$ \mujyb, $\sigma_{\rm GMRT} = 400$ \mujyb\ and $\sigma_{\rm VLA} = 70$ \mujyb. The negative $-3\sigma$ contours are shown in dashed. The beam sizes are $34\arcsec \times 23\arcsec$ (\lofar), $43\arcsec \times 29\arcsec$ (\gmrt) and $43\arcsec \times 38\arcsec$ (\vla) and are shown in the bottom left corners. More details on the images are reported in Tab.~\ref{tab:radio_imaging}.}
 \label{fig:radio_images}
\end{figure*}

\subsection{\lofar}\label{sec:lofar}

The \lotss\ observations are typically separated by $2.6\deg$ and we have analyzed the \lotss\ pointing that is centered closest to A1758 (offset by $\sim 1.1\deg$). The characteristics of this observation are summarized in Tab.~\ref{tab:radio_obs}. To calibrate the data we followed the facet calibration scheme described in \citet{vanweeren16calibration} and \citet{williams16}. This procedure comprises three steps, which we describe below. \\
\indent
In the first step we perform direction-independent calibration (\prefactor\footnote{https://github.com/lofar-astron/prefactor} pipeline). In this step, the flux calibrator (3C196) data are averaged and bad quality data are flagged. Complex gains and clock offsets for different antenna stations are calibrated off a 3C196 model adopting the absolute flux density scale of \citet{scaife12}. The amplitude and clock solutions are then transferred to the target data before an initial phase calibration against a sky model generated from the \tgssE-Alternative Data Release \citep[\tgss-ADR1;][]{intema17}. In the second step, wide field images covering the full \lofar\ field-of-view (FoV) are made from the products of the first step using \wsclean\ \citep{offringa14}. The \pybdsfE\ \citep[\pybdsf;][]{mohan15} software is then used to detect sources which are then subtracted from the \uv-data using the clean component models. Images are produced at medium ($\sim 40\arcsec \times 30\arcsec$) and low resolution ($\sim 120\arcsec \times 100\arcsec$) to ensure that both compact and extended sources are subtracted. The image sizes are about $12\deg \times 12\deg$ and $30\deg \times 30\deg$ for medium and low resolution, respectively. \\
\indent
In the direction-dependent calibration step (\factor\footnote{https://github.com/lofar-astron/factor} pipeline), nearly thermal noise limited images can be produced. It operates by using bright sources in the field to calibrate the phases and amplitudes in a restricted portion of the sky, \ie\ a ``facet''. This is needed as the \lofar\ primary beam is large, requiring many different corrections for ionospheric distortions and beam model errors across the FoV. For this reason, the FoV (usually a region within $2.5\deg$ from the pointing center) is divided into tens of facets that are typically processed in order of decreasing flux density of the calibrator source. After a number of self-calibration cycles on the facet calibrator, the fainter sources that were subtracted in the second step are added back into the data and are calibrated using the solutions derived from the facet calibrator. An updated sky model of the region covered by the facet is obtained by subtracting the components of the processed facet from the \uv-data. The \uv-data that is then used for the calibration of the subsequent facet has smaller systematics due to the source subtraction and the effective noise in the dataset is reduced. This process is iteratively repeated until all the desired directions have been calibrated.  This procedure has been successfully applied to image several other galaxy clusters with the \lofar\ HBA \citep[\eg][]{ vanweeren16toothbrush, degasperin17gentle, hoang17, wilber18a1132}. \\
\indent
All \lofar\ images in the paper are reported at the central observing frequency of 144~MHz and were produced with \casa\ v4.7 \citep{mcmullin07}. The imaging was done using the multi-scale multi-frequency deconvolution algorithm \citep[MS-MFS,][]{rau11}, with second order Taylor terms (\texttt{nterms = 2}), and W-projection \citep{cornwell05}. For the facet containing A1758, an image size of $5120\arcsec \times 5120\arcsec$ was adopted to ensure that all of the sources in the vicinity of the cluster were adequately deconvolved. An inner \uv-cut of $80\lambda$ (corresponding to an angular scale of $43\arcmin$) was also applied on the data to reduce sensitivity to very large scale emission from the Galaxy. Different resolution images were created using various different Briggs weightings \citep{briggs95} and by applying an \uv-taper, as reported in Tab.~\ref{tab:radio_imaging}. Primary beam correction was performed with \awimager\ \citep{tasse13}. Uncertainties in the flux scale that are caused by inaccuracies in the \lofar\ HBA beam model \citep[see][]{vanweeren16calibration, hardcastle16} were corrected by cross-matching a number of compact sources extracted from the \lofar\ image with the \tgss\ \citep{intema17}. Throughout the paper we have applied correction factor that was computed from the mean \lofar/\tgss\ integrated flux density ratio of 1.08 and a calibration error of 15\% on \lofar\ flux densities, which is in agreement with other \lofar\ HBA studies \citep[\eg][]{shimwell16, savini18}
 
\subsection{\gmrt}

We analyzed an 8~hr archival \gmrt\ 325~MHz observation of A1758 (details in Tab.~\ref{tab:radio_obs}). Data were reduced with the \spamE\ (\spam) package \citep{intema09}, which is an automated pipeline to process \gmrt\ observations based on the \aipsE\ (\aips). Here we outline the main steps of the \spam\ data reduction, for more details the reader is referred to \citet{intema09, intema17}. First, the dataset is averaged in time and frequency to reduce the data processing time whilst keeping enough resolution in both time and frequency to avoid smearing. Bad data due to corrupted baselines, non-working antennas, and radio frequency interference were also excised. The bandpass and absolute flux density scale were calibrated using the primary flux calibrator 3C147 and adopting the \citet{scaife12} flux scale. An initial phase-only calibration using a sky model generated from the \vlssE\ Redux \citep[\vlss r;][]{lane14}, the \wenssE\ \citep[\wenss;][]{rengelink97} and the \nvssE\ \citep[\nvss;][]{condon98} was followed by a number of loops of self-calibration, wide-field imaging and additional flagging of bad data. Then the bright sources in the primary beam are used to perform a direction-dependent calibration and ionospheric modeling aiming to mitigate the phase errors introduced by the ionosphere. The final calibrated data were then imaged with \casa\ v4.7, in the manner that was described at the end of Section~\ref{sec:lofar}. In our analysis we did not consider the effect of variation in system temperature \citep[see][]{sirothia09}. Instead, we adopted a similar approach to that described in Section~\ref{sec:lofar}, cross-matching a number of sources extracted from the \gmrt\ image with the \wenss\ \citep{rengelink97} and applying a correction factor of 0.73 on the \gmrt\ flux densities with a systematic uncertainty of 15\% \citep[see][]{chandra04}.


\subsection{\vla}

We analyzed archival \vla\ observations of A1758 at 1.4~GHz in configurations C and D. The details of the observations are reported in Tab.~\ref{tab:radio_obs}. Data reduction was performed with \aips\ where the two datasets were edited, calibrated and imaged separately. Thus the \uv-data were combined to produce a single image of the cluster. The flux calibrator of \vla\ observations was 3C286 \citep[model from][]{perley13}. The final imaging was performed with \casa\ v4.7 as described in Section~\ref{sec:lofar}. The absolute flux scale calibration errors were conservatively set to 5\% on \vla\ flux densities.

\subsection{Integrated synchrotron spectra and source subtraction}\label{sec:alpha}

Given the different \uv-coverage of the \lofar, \gmrt\ and \vla\ observations, it was necessary to match the \uv-sample of the different interferometers as closely as possible to provide an accurate comparison between the flux densities measured at different frequencies, and to compute the diffuse emission spectra. \\
As a first step, we removed the discrete sources from each datasets. This procedure was performed by applying an inner \uv-cut of $2.0$ k$\lambda$ (corresponding to an angular scale of $103\arcsec$, \ie\ about 440 kpc at $z=0.279$) to the data to image only the compact sources in the field whose clean components were subsequently subtracted from the visibilities. To image the diffuse emission after the source subtraction, for each dataset we used the an inner \uv-cut of $170\lambda$ and \texttt{uniform} weighting. A Gaussian \uv-taper of $35\arcsec$ was also used to enhance diffuse emission and to produce images with comparable beams. Errors on flux densities were estimated via

\begin{equation}\label{eq:flux_error}
 \Delta S = \sqrt{\left( \sigma_{rms} \times \sqrt{\frac{A_s}{A_b}} \right)^2 + (\xi \times S)^2}
\end{equation}

\noindent
where $\xi$ is the calibration uncertainty, $\sigma_{rms}$ is the root-mean-square noise while $A_s$ and $A_b$ are the source and beam areas, respectively. \\
\indent
In Fig.~\ref{fig:radio_images} we present the \lofar\ image (left panel) along with the point-source-subtracted images from the \gmrt\ (central panel) and \vla\ (right panel). This choice has been made since the diffuse emission is best visible.

\subsection{\chandra}\label{sec:chandra}

\begin{figure}
 \centering
 \includegraphics[width=.73\hsize,angle=-90]{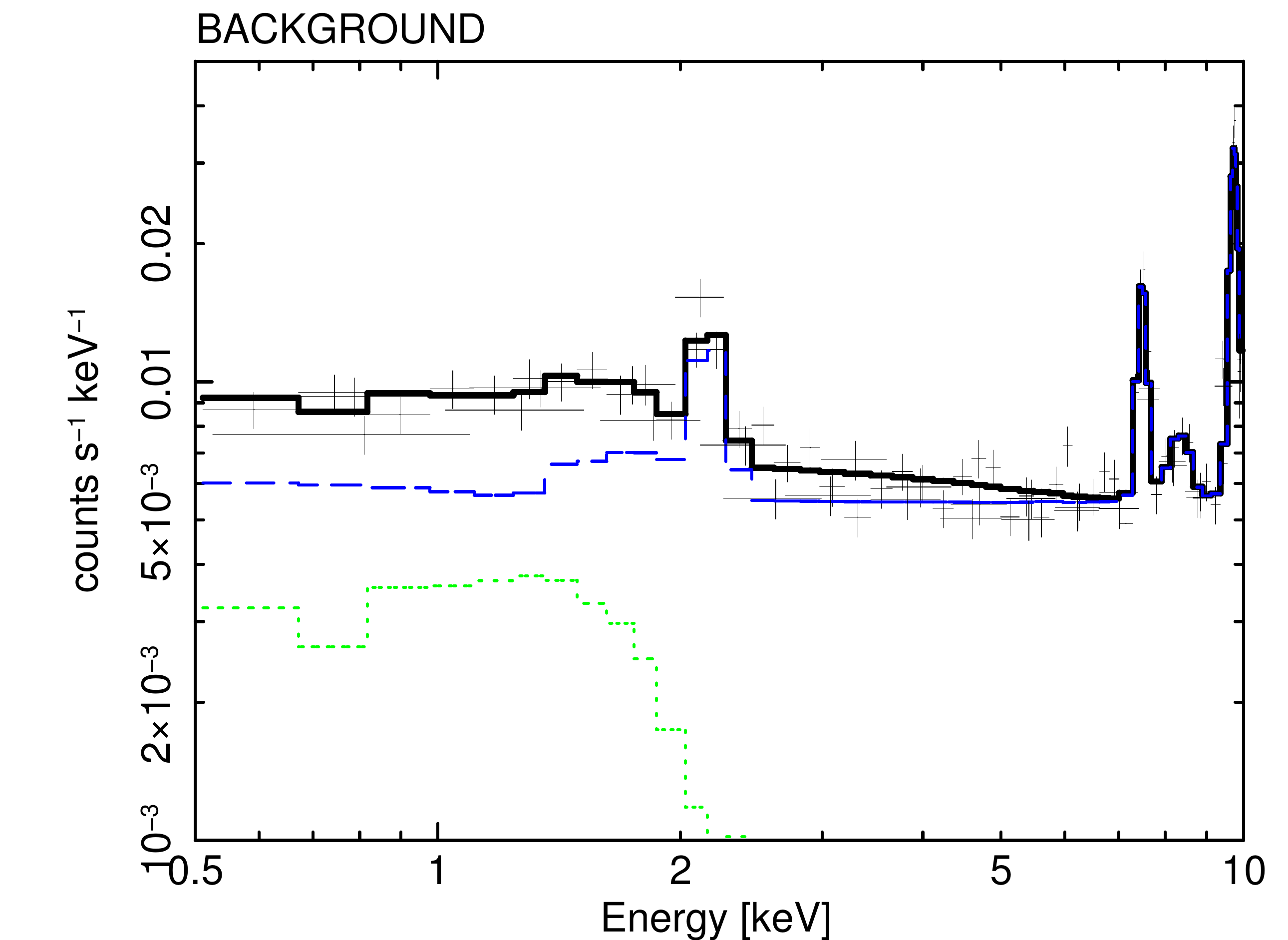}
 \caption{Spectrum of the \chandra\ background. Data points are shown in black together with the best-fitting model. The astrophysical and instrumental backgrounds are shown in dotted green and dashed blue, respectively. Whilst the three \obsid\ spectra were simultaneously fitted, the models for only one observation were reported in order to avoid confusion in the plot. The \cstatdof\ of the fit is 406/386.}
 \label{fig:background}
\end{figure}

We retrieved three \acisi\ observations (\obsid: 13997, 15538, 15540) on A1758 from the  \chandra\ data archive\footnote{http://cda.harvard.edu/chaser/} for a total exposure time of 150 ks. We mention that two other \chandra\ pointings on A1758 also exist; however they composed of an \aciss\ observation where only A1758N is in the FoV (\obsid\ 2213) and by a short (7 ks) observation (\obsid\ 7710) whose exposure time is negligible with respect to the total integration time. For that reason they were not used in this work. \\
\indent
Data reduction was performed with \ciao\ v4.9 and \chandra\ \caldb\ v4.7.3. Time periods affected by soft proton flares were removed by inspecting the light curves in the $0.5-7.0$ \kev\ band extracted for each \obsid\ from the S2 chip with the \texttt{lc\_clean} routine. After this step, the resulting clean exposure time is 137 ks. We used \texttt{merge\_obs} to add together the three datasets and produce the final cluster image in the $0.5-2.0$ \kev\ band. An exposure-corrected point spread function (PSF) map with minimum size was created from the combination of the PSF and exposure maps of the three \obsid s. This was used to detect discrete sources with the \texttt{wavdetect} task, which were later confirmed by eye and excluded in the further analysis. \\
\indent
Spectra were extracted in the same regions from all the \obsid s and simultaneously fitted in the $0.5-10.0$ \kev\ band with \xspec\ v12.9.0o \citep{arnaud96}. The background was carefully treated with a model that included both astrophysical and instrumental emission components, as shown in Fig.~\ref{fig:background}. The former is described by two main components due to: the Galactic emission, modeled with two thermal plasmas with $kT_1 = 0.14$ \kev\ and $kT_2 = 0.25$ \kev, and the cosmic X-ray background, described with an absorbed power-law with photon index $\Gamma=1.4$. For the latter we followed \citet{bartalucci14} which provided an analytical model for the \acisi\ particle background. Spectra were fitted adopting Cash statistics \citep{cash79} and an absorbed thermal model for the ICM with metallicity fixed at 0.3 \zsun\ (solar abundance  table by \citealt{anders89}) and hydrogen column density $\nh = 1.03 \times 10^{20}$ cm$^{-2}$ computed from the Leiden/Argentine/Bonn Survey of Galactic H\textsc{i} \citep{kalberla05}. \\
\indent
We used \contbin\ v1.4 \citep{sanders06contbin} to compute the thermodynamical properties of the ICM in A1758. A signal-to-noise ratio (S/N) of 50 for the net counts in the $0.5-2.0$ \kev\ band was set to delineate the regions where spectra were extracted and fitted as written above. For more details on the computation of the maps of the ICM thermodynamical quantities and generally on the \chandra\ data analysis we refer the reader to \citet{botteon18edges} in which the same procedures adopted in the current paper have been more thoroughly described.

\begin{figure}
 \centering
 \includegraphics[width=\hsize]{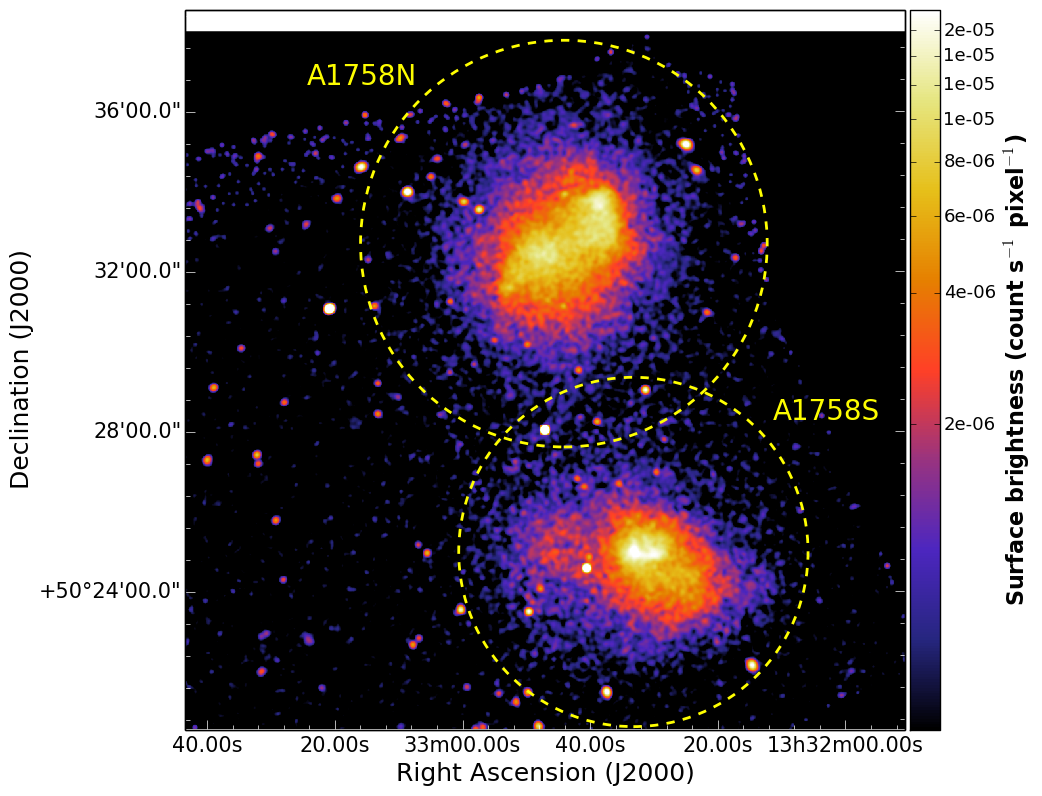}
 \caption{\chandra\ exposure-corrected image in the $0.5-2.0$ \kev\ band of A1758 smoothed to a resolution of $\sim3\arcsec$. Yellow circles indicate the approximate location of \rfive\ for each cluster.}
 \label{fig:chandra_only}
\end{figure}

\section{Results}

\subsection{A1758N radio halo}

\begin{figure*}
 \centering
 \includegraphics[width=.95\textwidth]{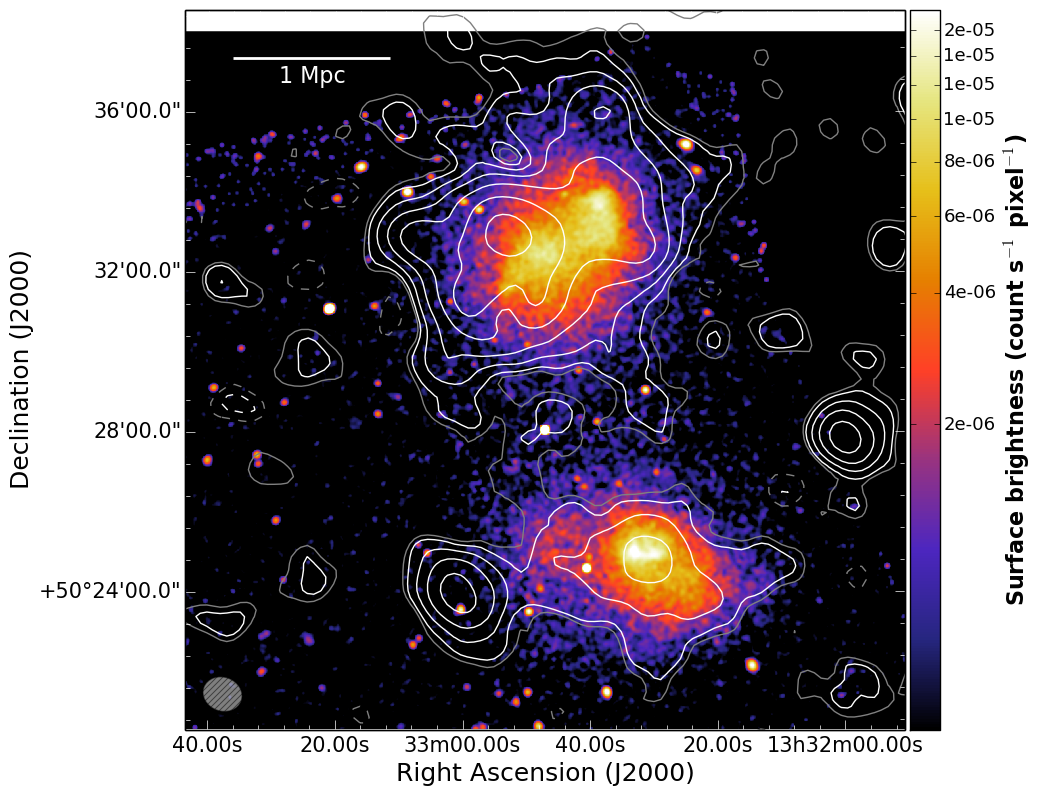}
 \caption{\lofar\ radio contours with point-sources subtracted of A1758 overlaid on the \chandra\ color image of Fig.~\ref{fig:chandra_only}. The \lofar\ white contours are spaced by a factor of 2 starting from $3\sigma$, where $\sigma_{\rm LOFAR} = 390$ \mujyb. The negative $-3\sigma$ contours are shown in dashed. Gray contours correspond to the $\pm2\sigma$ level. The beam size is $60\arcsec \times 51\arcsec$ and is shown in the bottom left corner. More details on the \lofar\ image are reported in Tab.~\ref{tab:radio_imaging}.}
 \label{fig:chandra_lofar_lowres}
\end{figure*}

\begin{figure*}
 \centering
 \includegraphics[width=.44\textwidth]{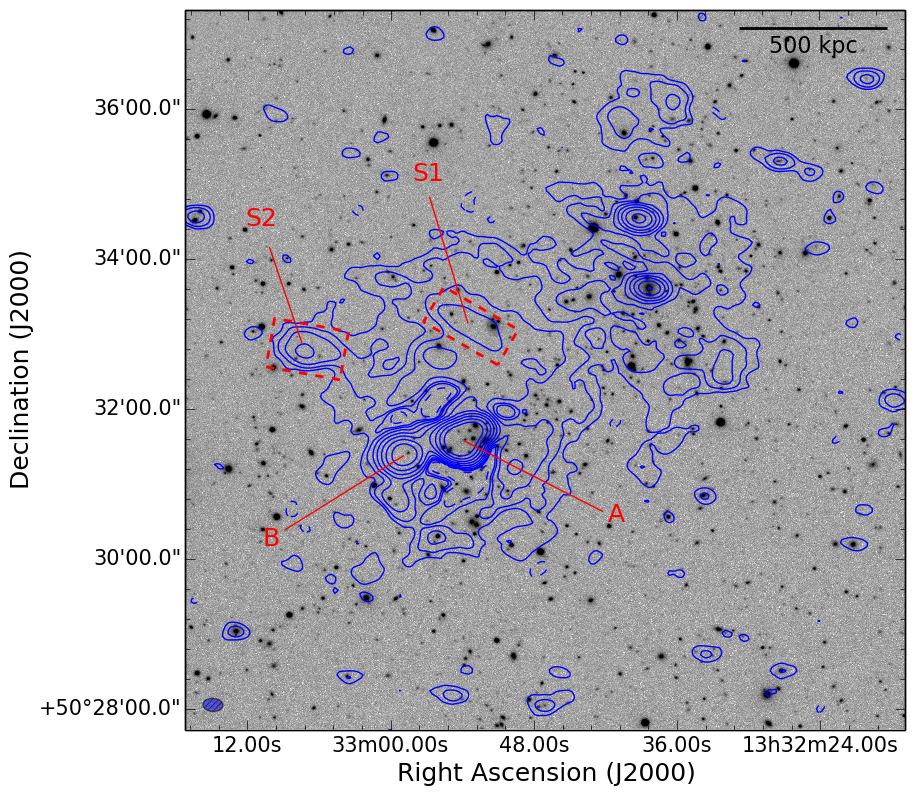} 
 \hspace{0.5cm}
 \includegraphics[width=.463\textwidth]{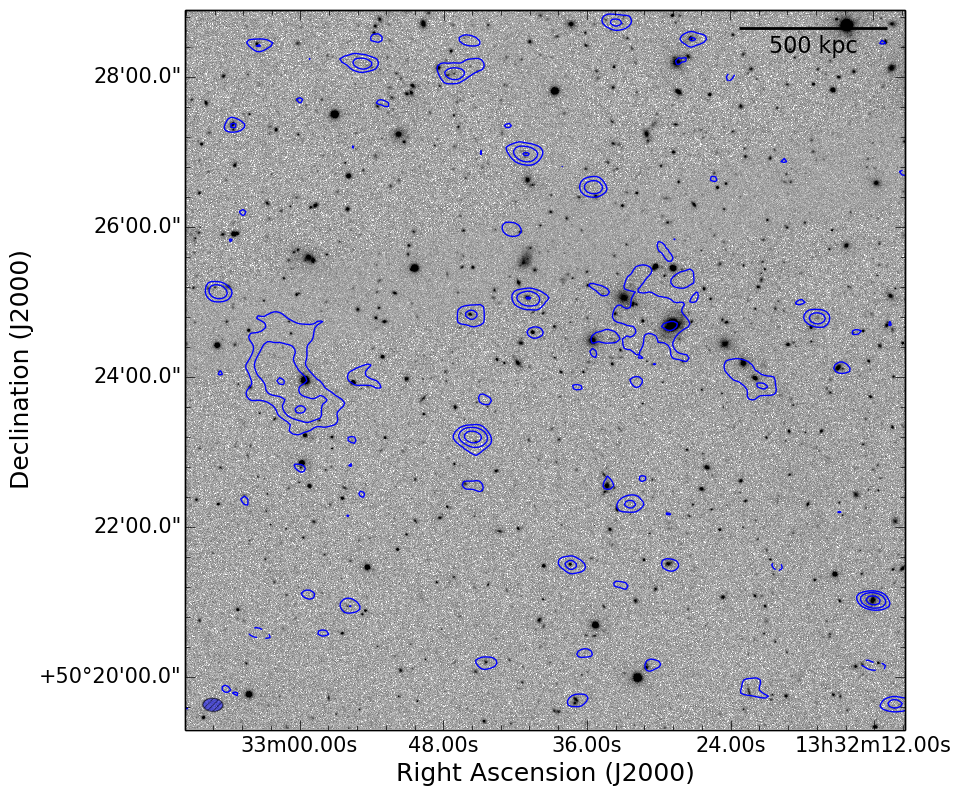}
 \caption{Mosaic \sdss g,r,i images of A1758N (\textit{left}) and A1758S (\textit{right}) overlaid with the \lofar\ contours spaced by a factor of 2 starting from $3\sigma$, where $\sigma_{\rm LOFAR} = 140$ \mujyb. The negative $-3\sigma$ contours are shown in dashed. The beam size is $16\arcsec \times 11\arcsec$ and is shown in the bottom left corners. At this resolution the radio halo in A1758S is marginally visible. More details on the \lofar\ image are reported in Tab.~\ref{tab:radio_imaging}.}
 \label{fig:sdss_lofar}
\end{figure*}

\begin{figure}
 \centering
 \includegraphics[width=\hsize]{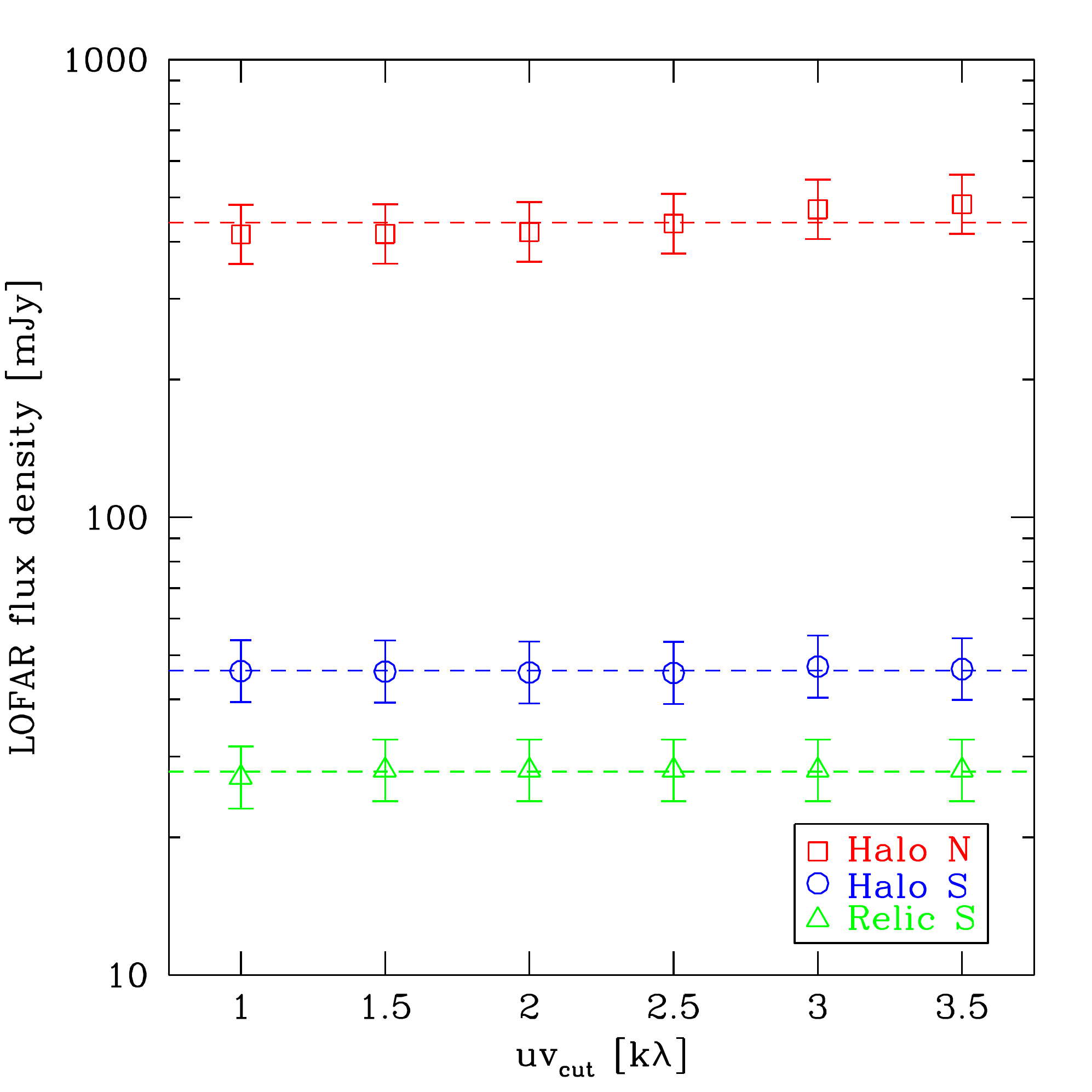}
 \caption{The flux densities of the diffuse emission in A1758 measured with \lofar\ versus the inner \uv-cuts adopted to subtract the point-sources. Dashed horizontal lines show the mean values of the measurements.}
 \label{fig:uvcut}
\end{figure}

\begin{figure}
 \centering
 \includegraphics[width=\hsize]{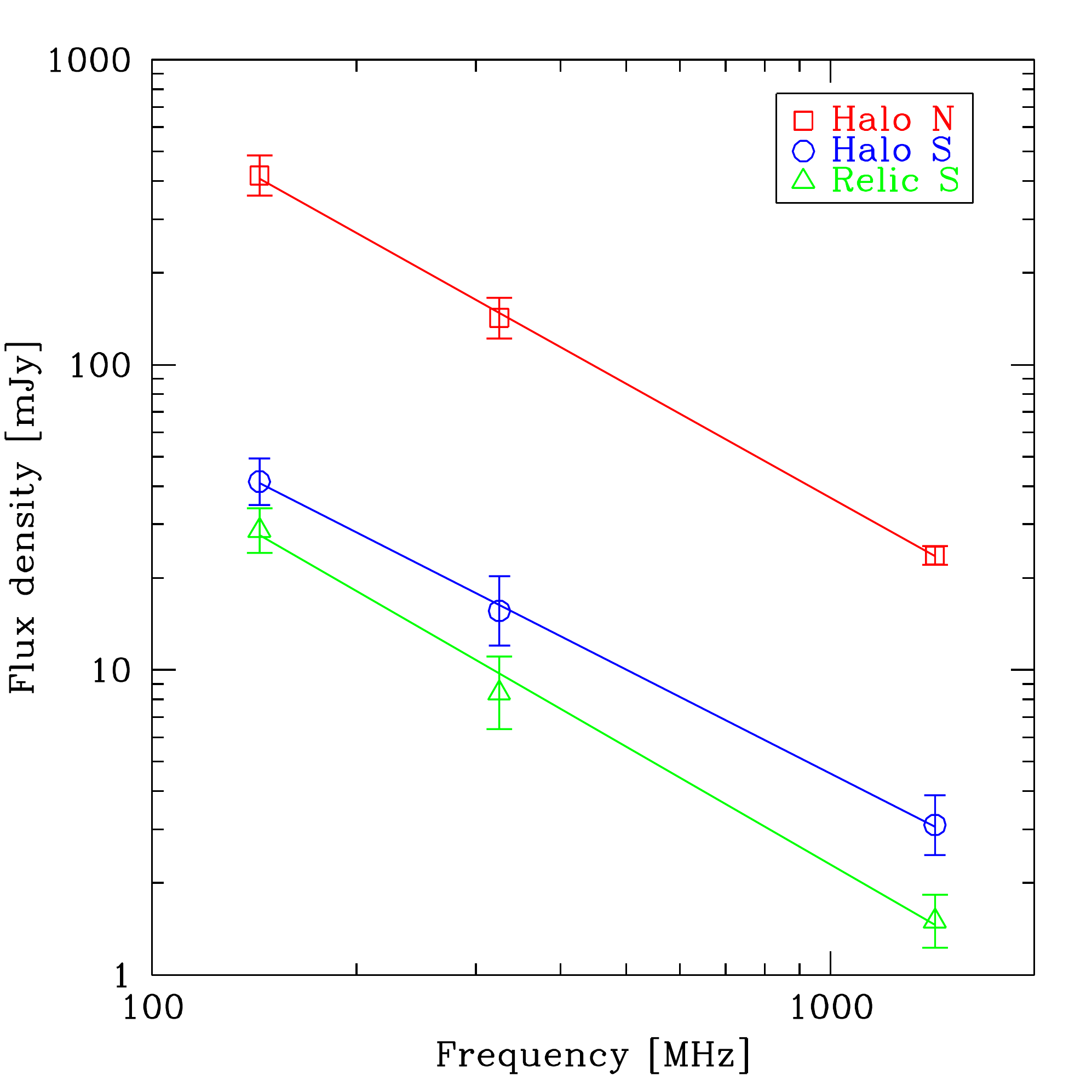}
 \caption{Integrated spectra of the diffuse radio emissions in A1758. The spectral index values are reported in Tab.~\ref{tab:fluxes}.}
 \label{fig:spectra_fit}
\end{figure}

\begin{table}
 \centering
 \caption{Flux densities of the diffuse emission in A1758. The spectral indexes were computed adopting the procedure described in Section~\ref{sec:alpha}.}
 \label{tab:fluxes}
  \begin{tabular}{lrrr} 
  \hline
  $\nu$ [MHz] & \multicolumn{3}{c}{$S_\nu$ [mJy]}  \\
  & Halo N & Halo S & Relic S \\
  \hline
  144 & $420\pm63$ & $45.8\pm7.1$ & $28.0\pm4.3$ \\
  325 & $134\pm20$ & $16.8\pm3.5$ & $8.9\pm2.0$ \\
  1425 & $24.7\pm1.7^\dagger$ & $3.1\pm0.7^\dagger$ & $1.5\pm0.3$ \\
  \hline
  $\alpha$ & $1.2\pm0.1$ & $1.1\pm0.1$ & $1.3\pm0.1$ \\
  \hline
  \multicolumn{4}{{p{.36\textwidth}}}{\textit{Notes.} $^\dagger$ The error takes into account also the uncertainties of the source subtraction.}
  \end{tabular}
\end{table}

Diffuse emission in A1758N is visible both from the \nvss\ and the \wenss\ surveys \citep{kempner01}. The observations taken with the \vla\ at 1.4 GHz \citepalias{giovannini09} and with the \gmrt\ at 325 MHz \citepalias{venturi13} confirmed the presence of a giant radio halo which is elongated in the NW-SE direction and only partially covering the X-ray emission of the cluster. The spectral index reported between these two frequencies is $\alpha=1.31\pm0.16$ \citepalias{venturi13}. \\
\indent
\lofar\ detects the extended radio halo flux in A1758N at higher significance and the recovered morphology appears consistent with the \gmrt\ and \vla\ maps, as demonstrated in Fig.~\ref{fig:radio_images}. We measure a largest linear size of the emission of $\sim2.2$ Mpc. The low resolution point-source-subtracted \lofar\ contours displayed in Fig.~\ref{fig:chandra_lofar_lowres} suggest that the non-thermal radio emission in A1758N covers the X-ray bright region of the cluster. At higher resolution (Fig.~\ref{fig:sdss_lofar}, left panel), only the brightest part of the radio halo is visible; in particular, \lofar\ shows two bright and straight structures (labeled as S1 and S2 in Fig.~\ref{fig:sdss_lofar}, left panel) apparently not associated with any optical galaxy. They might indicate regions where the plasma has been somehow locally compressed or re-accelerated \citep[\eg][]{shimwell16, degasperin17gentle}. The feature S1 is also detected with the \gmrt\ and \vla\ (Fig.~\ref{fig:radio_images}). In the southeast, A1758N also hosts two prominent narrow angle tailed radio galaxies labeled as A and B in Fig.~\ref{fig:sdss_lofar} (left panel). The former (also identified as $1330+507$) was studied at high resolution with the \vla\ by \citet{odea85}. \\
\indent
The presence of resolved radio galaxies embedded in the halo (\eg\ source A and B in Fig.~\ref{fig:sdss_lofar}, left panel) makes it difficult to disentangle their contribution from that of the halo. We repeated the subtraction by adopting inner \uv-cuts in the range $1.0-3.5$ k$\lambda$, corresponding to linear sizes of $873-249$ kpc at the cluster redshift, to assess the uncertainties in our source subtraction on the \lofar\ dataset, in addition to the procedure described in Section~\ref{sec:alpha}. In Fig.~\ref{fig:uvcut} we show how the flux density measurement of the northern radio halo varies with the \uv-cut, ranging from 415 to 483 m\jy\ (the mean value is 440 m\jy). This indicates that the choice of the \uv-cut has an impact on the halo in A1758N. In contrast, the integrated flux density of the diffuse sources in A1758S (see Sections below) is essentially independent on the \uv-cut used (Fig.~\ref{fig:uvcut}), indicating that the subtraction is less problematic which is expected as there are just a few weak discrete sources without significant extended emission (\cf\ Fig.~\ref{fig:sdss_lofar}, right panel). \\
\indent
In Tab.~\ref{tab:fluxes} and Fig.~\ref{fig:spectra_fit} we report the flux density measurements at three frequencies and the spectra, respectively, of the diffuse emission in A1758. The spectral index between 144~MHz and 1.4~GHz is $\alpha=1.2 \pm 0.1$ for the halo in A1758N, consistent with that of $\alpha=1.31\pm 0.16$ computed by \citetalias{venturi13}. The flux densities measured in our \gmrt\ and \vla\ images agree to that reported in \citetalias{venturi13} within $1\sigma$. The emission from the potential relic and the halo in A1758S have not been previously reported.

\begin{figure*}
 \centering
 \includegraphics[width=.33\textwidth]{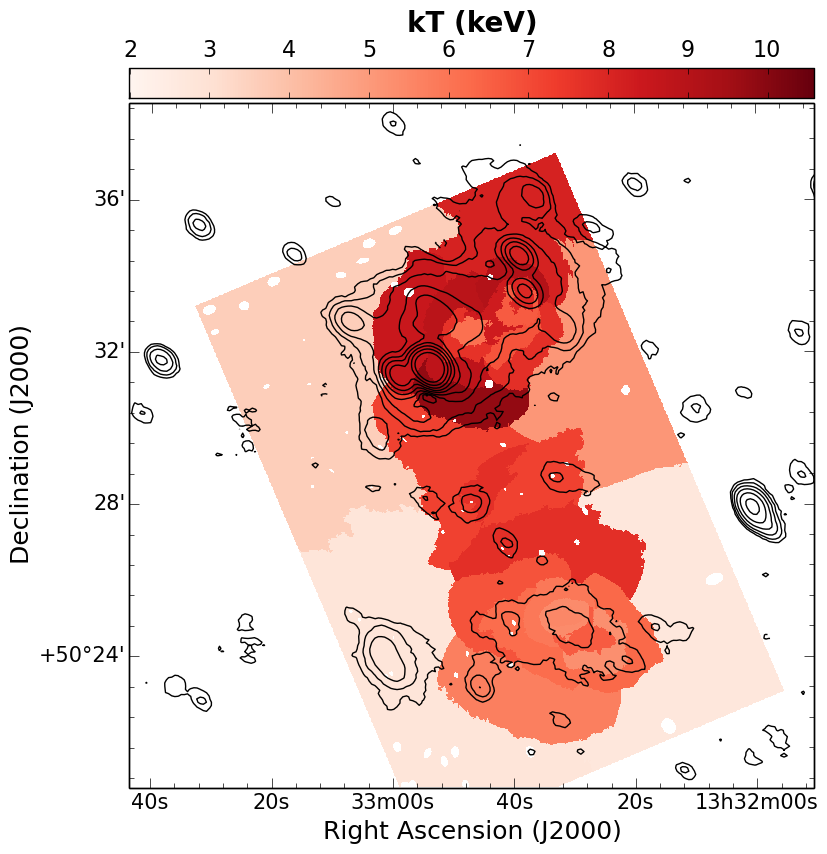}
 \includegraphics[width=.33\textwidth]{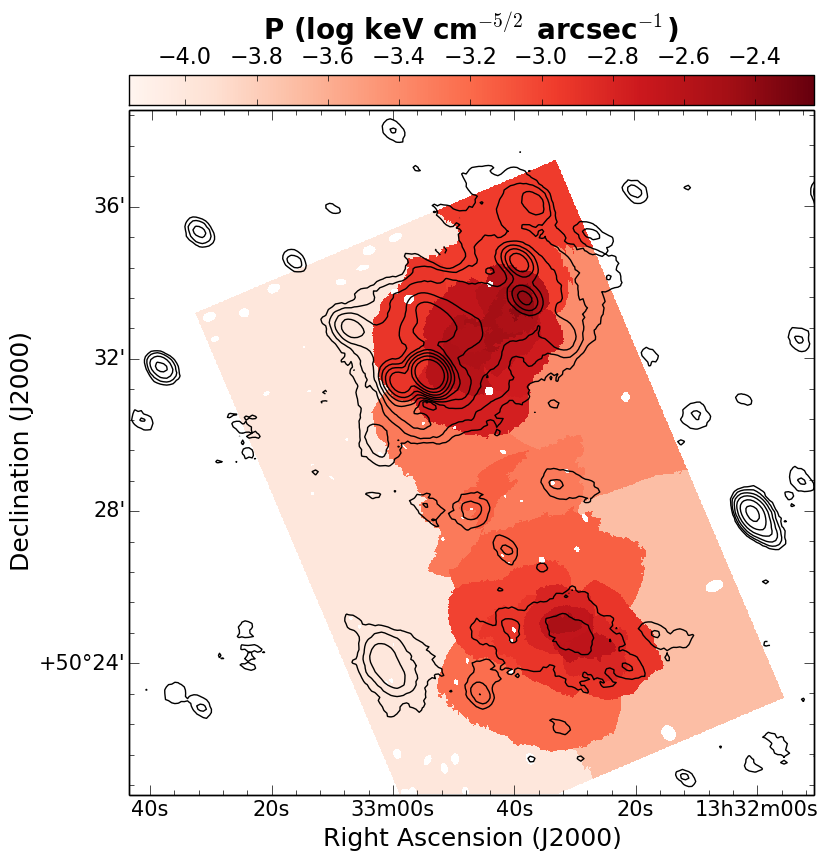}
 \includegraphics[width=.33\textwidth]{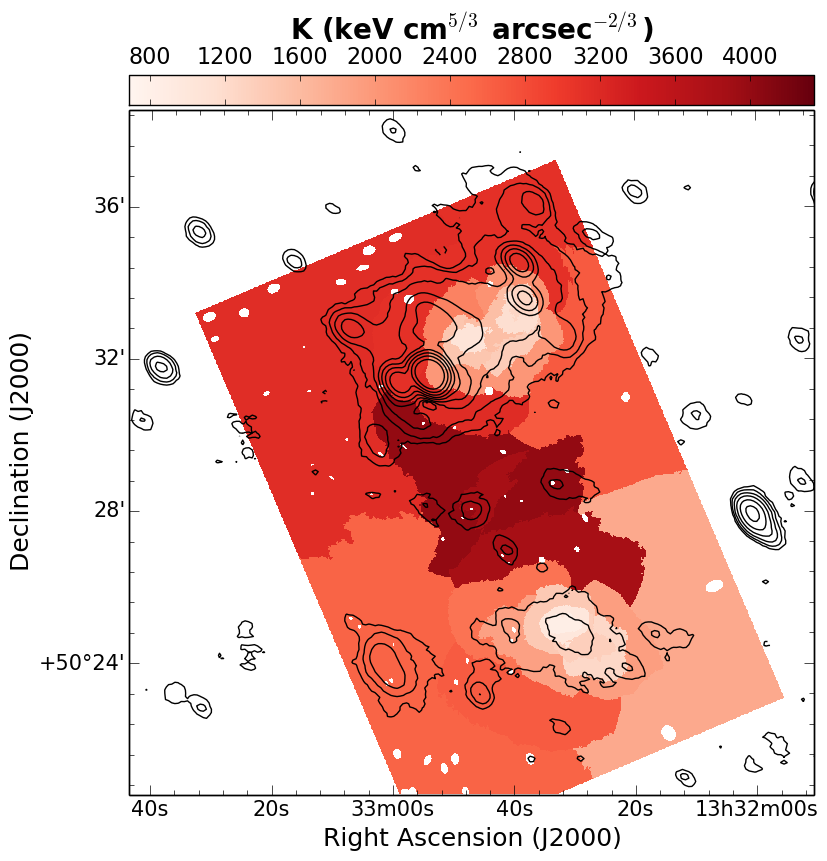}
\caption{Thermodynamical properties of the ICM in A1758 with overlaid the \lofar\ contours of Fig.~\ref{fig:radio_images}. Images depict projected values of temperature (\textit{left}), pressure (\textit{center}) and entropy (\textit{right}). The temperature error map and a lower S/N temperature map are reported in Appendix~\ref{app:errormap}.}
 \label{fig:icm_maps}
\end{figure*}

\subsection{A1758S radio halo}

The halo in A1758S is barely visible with the \gmrt\ and it is detected at low significance with the \vla\ (Fig.~\ref{fig:radio_images}). The characteristics of the emission recovered by the \lofar\ observation (Fig.~\ref{fig:chandra_lofar_lowres}) are typical for a radio halo, \ie\ low surface brightness, similar morphology with respect to the ICM thermal emission and a largest linear size of $\sim1.6$ Mpc. It is worth noting that the merger axis of A1758S is likely close to the line of sight \citep[\eg][]{monteirooliveira17a1758}, hence we can not fully discard the possibility that the radio emission traces a radio relic observed face-on, although this is unlikely because the remarkable similarity between the radio and the X-ray emission (Fig.~\ref{fig:chandra_lofar_lowres}). Future studies on the source polarization level will definitely clarify this point. \\
\indent
The integrated flux density measured with \lofar\ within the $3\sigma$ contour from the low resolution image of Fig.~\ref{fig:chandra_lofar_lowres} (excluding the peripheral emission to the east, see Section below) is $45.8\pm7.1$ m\jy, \ie\ one order of magnitude lower than that of the halo in A1758N.  Indication of the presence of a radio halo in A1758S has also been found in the \wsrtE\ (\wsrt) data at 367~MHz \citep{drabent17}. However, the available observation is not suitable to study in detail the diffuse emission due to its inadequate angular resolution which makes the point-source subtraction unreliable. The spectral index computed between 144~MHz and 1.4~GHz within a region traced by the \lofar\ emission is $\alpha=1.1 \pm 0.1$ (see Tab.~\ref{tab:fluxes}) and the fit is shown in Fig.~\ref{fig:spectra_fit}. We also determined the halo spectral index considering a region defined by the \vla\ $3\sigma$ contour. In this case, the flux densities evaluated in the \lofar\ and \vla\ images are $\sim12.5$ m\jy\ and $\sim1.2$ m\jy, respectively, and the spectral index is consistent with that reported above.

\subsection{A1758S candidate radio relic}

To the east of A1758S, an extended radio source at the boundaries of the X-ray emission is observed with \lofar, \gmrt\ and \vla\ (Fig.~\ref{fig:radio_images}). In the \lofar\ low resolution contours of Fig.~\ref{fig:chandra_lofar_lowres}, the $3\sigma$ contour of this emission is connected with that of the radio halo. In Tab.~\ref{tab:fluxes} we report the flux density measurements at various frequencies. We estimated a spectral index between 144~MHz and 1.4~GHz of $\alpha = 1.3 \pm 0.1$ for this source (Fig.~\ref{fig:spectra_fit}). We tentatively classify this emission as a radio relic based on the following characteristics: (i) its elongated morphology roughly arc-shaped and perpendicular to the thermal cluster emission, (ii) its largest linear size $>500$ kpc, (iii) its peripheral location in the same direction of the ICM elongation, (iv) its steep spectrum, and (v) the absence of a clear optical counterpart\footnote{Note that the source roughly at the center of the diffuse radio emission in the optical image of Fig.~\ref{fig:sdss_lofar} (right panel) is a star. No redshift has been reported for the X-ray point source (identified as \sdss\ J133300.32+502332.2) embedded in the candidate relic that is visible in the \chandra\ image of Fig.~\ref{fig:chandra_lofar_lowres}. With the current data we can not conclude whether it is associated with the radio emission.} and/or bright compact radio emission (Fig.~\ref{fig:sdss_lofar}, right panel). All these properties are commonly observed in radio relics but can also be seen in other objects, such as dead radio galaxies \citep[\eg][and references therein]{brienza16}. A definitive claim would require either the study of the spectral index gradient toward the cluster center, measurements of the source polarization or the detection in the X-rays of an underlying shock front. Unfortunately none of these measurements can be carried out with the present data. 

\subsection{X-ray properties of A1758N and A1758S}

\begin{figure*}
 \centering
 \includegraphics[width=.5\textwidth,trim={4.6cm 1.6cm 3.4cm 0cm},clip,valign=c]{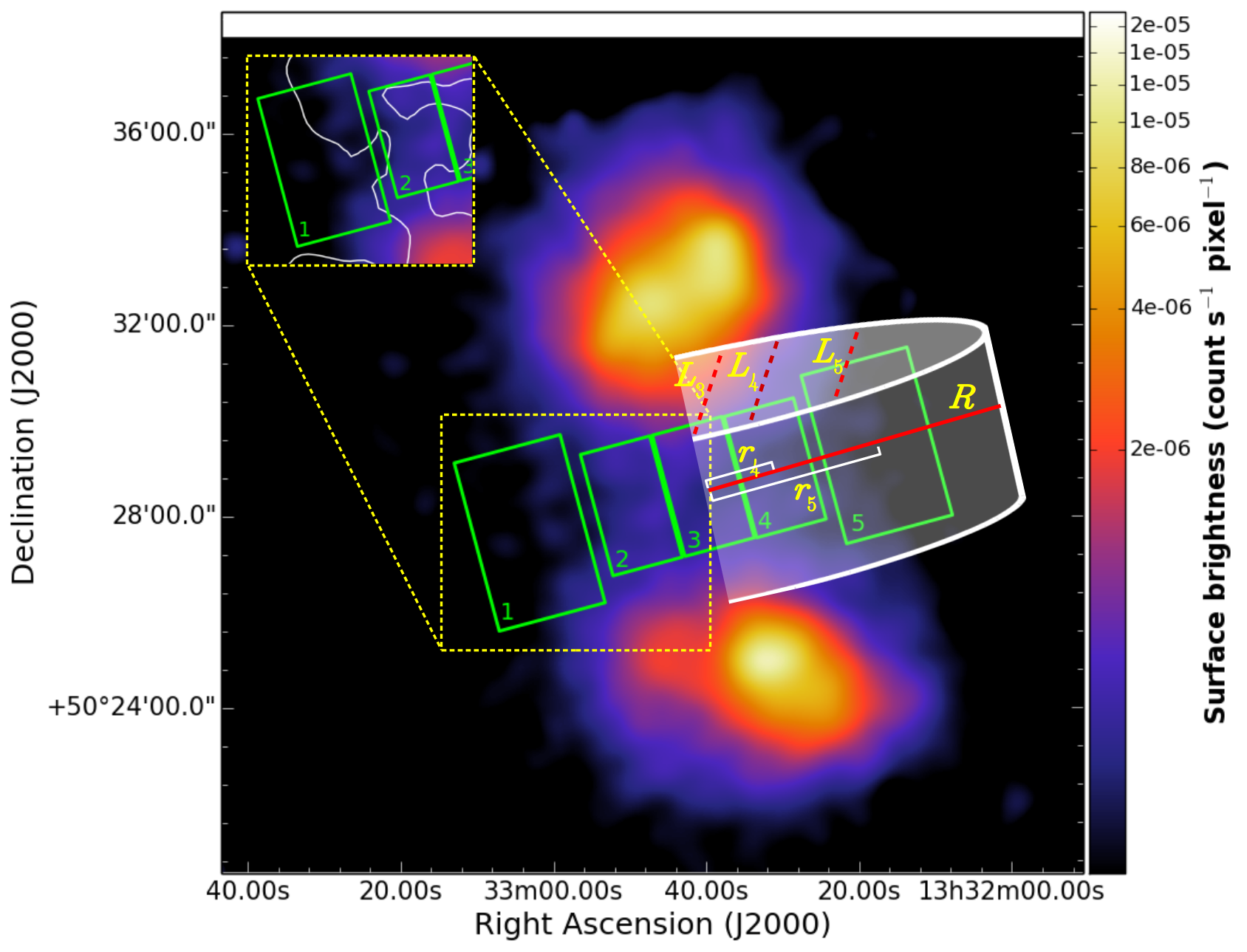} 
 \hspace{0.5cm}
 \includegraphics[width=.45\textwidth,trim={0cm 0.2cm 1cm 0.8cm},clip,valign=c]{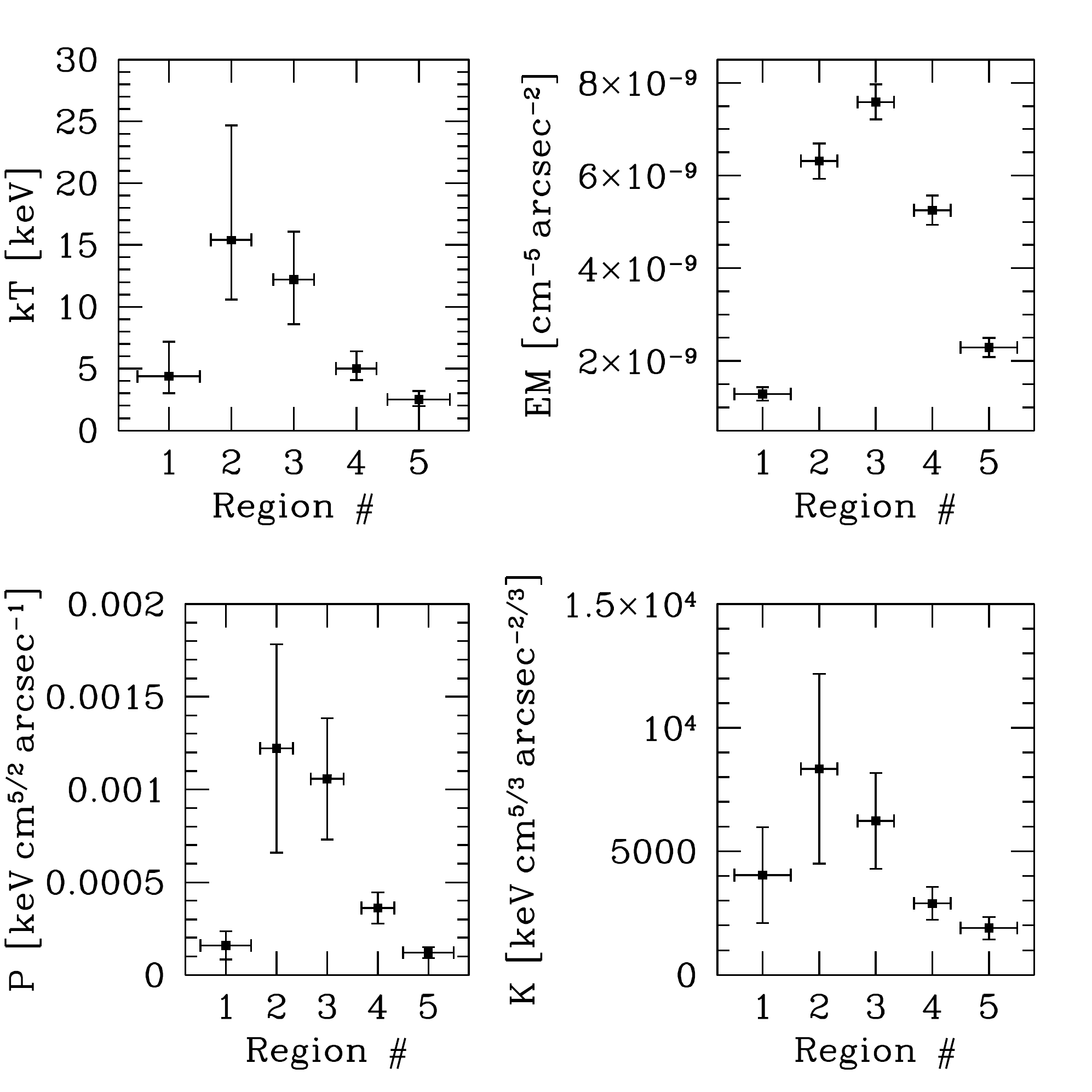}
 \caption{\textit{Left panel}: same \chandra\ image of Fig.~\ref{fig:chandra_only} but with point-sources subtracted and smoothed to a resolution of $\sim15\arcsec$ to highlight the X-ray channel between the clusters. The spectral extracting regions are overlaid in green. The inset panel shows the spatial coincidence between the tentative radio bridge and the putative post-shock region. The cylindrical model assumed to asses the effects of projection is also sketched. The line of sight across the $i$-region is $L_{i} \sim 2 \sqrt{R^2 - r_{i}^{2}}$. \textit{Right panels}: projected values of temperature ($kT$), emission measure ($EM$), pressure ($P$) and entropy ($K$) of the X-ray channel.}
 \label{fig:shock}
\end{figure*}

The deep \chandra\ observation of A1758 allowed us to derive the projected maps of the ICM thermodynamical quantities shown in Fig.~\ref{fig:icm_maps}. The temperature map displays overall higher values in A1758N than in A1758S. These values are within the range of $8.0-9.9$ \kev, for A1758N, and $6.0-6.7$ \kev, for A1758S, that were reported by \citet{david04} who made measurements within 1~Mpc radius aperture centered on the centroid of each cluster. Shock heated regions toward the NW and SE of the northern cluster are suggested by high values of temperature and pressure (Fig.~\ref{fig:icm_maps}, see also Fig.~\ref{fig:low_snr_kt}), in agreement with the late merger scenario \citep{david04} where the shocks have already crossed the central region of the ICM and are moving outwards with high Mach numbers \citep{machado15a1758}. Our entropy map of Fig.~\ref{fig:icm_maps} (right panel) highlights the presence of the two cores in A1758N and the single core in A1758S, characterized by the lowest values of entropy in the map, in line with the bimodal (A1758N) and single clump (A1758S) mass distribution already inferred from optical studies \citep{dahle02, okabe08, ragozzine12, monteirooliveira17a1758}. 

\subsection{The bridge between A1758N and A1758S}

The maps of the ICM thermodynamical quantities shown in Fig.~\ref{fig:icm_maps} show a complex thermodynamics in the region between A1758N and A1758S, suggesting that the two clusters are in early interaction. At this stage of the interaction the gas between them is compressed and heated, somewhat explaining the observed high values of temperature ($\sim 7.5$ \kev) and pressure in the region between the cluster pair. Moreover, the entropy map (Fig.~\ref{fig:icm_maps}, right panel) displays the largest values in such a region, further indicating an unrelaxed state of the clusters outskirts. \\
\indent
The \lofar\ low resolution contours of Fig.~\ref{fig:chandra_lofar_lowres} give a tantalizing hint of a low surface brightness bridge connecting A1758N and A1758S. This emission is detected at the $2\sigma$ level towards the eastern edge of the region between the two clusters. On the western edge of this region, a protuberance of the A1758N halo extends towards A1758S. Although particular care was devoted in the subtraction of the point-sources between the clusters, the blending of low level residual emission due to faint and unresolved sources, combined with the large synthesized beam of the image and with the non-uniform distribution of the noise, could mimic the filamentary structure. All this, together with the low significance level of the emission, does not allow us to make a firm statement about its presence. Nonetheless, filaments connecting galaxy clusters are expected to be observed in the radio band even on larger scales \citep[\eg][]{keshet04imprint, arayamelo12, vazza15forecasts}. We note that a hint of a Sunyaev-Zel'dovich (SZ) signal connecting A1758N and A1758S was reported also in \citet{ami12}. \\
\indent
We analyzed the thermal properties of the X-ray emission between A1758N and A1758S, visible by smoothing to a resolution of $\sim15\arcsec$ the \chandra\ $0.5-2.0$ \kev\ image (Fig.~\ref{fig:shock}, left panel). We extracted spectra from five regions enclosing $\sim1000$ counts each in the $0.5-2.0$ \kev\ band that were fitted as described in Section~\ref{sec:chandra}. The best-fitting spectra are reported in Appendix~\ref{app:spectra}. These were used to compute the temperature profile shown in Fig.~\ref{fig:shock} (right panel). We measure high $kT$ values inside the X-ray channel that drop by a factor of $\sim3$ between regions 2 and 1 (see Appendix~\ref{app:errormap} for the apparent discrepancy between the values of the temperature map and profile shown in Fig.~\ref{fig:shock}); if projection effects play a role, the temperature drop would be even larger. Unfortunately, the count statistics does not allow us to increase spatial resolution to firmly understand if this is sharp drop or a gradual decrement. If we assume that this is a jump due to a shock and we apply the Rankine-Hugoniot jump conditions \citep[\eg][]{landau59}

\begin{equation}\label{eq:mach-from-temp}
 \frac{T_2}{T_1} = \frac{5\mach_{\rm kT}^4 + 14\mach_{\rm kT}^2 -3}{16\mach_{\rm kT}^2}
\end{equation}

\noindent
the derived Mach number\footnote{Although also the surface brightness drops outside the X-ray channel, we did not attempt the ``canonical'' broken power-law density profile fit \citep[\eg][]{markevitch07rev} to search for the edge due to the complex geometrical problem at this location (overlap of the outskirts of two galaxy clusters).} would be $\machkt = 3.0^{+1.4}_{-1.0}$. We notice that this putative shock is co-located with the $2\sigma$ level emission observed by \lofar\ (Fig.~\ref{fig:shock}, inset in the left panel). This is tantalizing and deserves future follow-ups as it might suggest a connection between the shock and the possible radio bridge. Its uncommonly high Mach number and unusual transversal location are in agreement with the recent work of \citet{ha18}, where these kind of shocks are referred as ``equatorial''. Alternatively, the high $kT$ values of the X-ray channel could be due to the adiabatic compression of the gas in the filament connecting A1758N and A1758S during the initial stage of the merger. \\
\indent
As a complementary information, the five spectra were used to compute the profiles of emission measure, pressure and entropy shown in Fig.~\ref{fig:shock} (right panel). These quantities are also observed to jump from the external to internal regions. We urge caution when interpreting these measurements as they were derived from the normalization of the cluster thermal model, that is $\propto n^2 L$, and are projected along the line of sight $L$ ($n$ is the density of the medium). For this reason they are usually referred as ``pseudo'' quantities \citep[\eg][]{mazzotta04}. We can asses the effects of projection assuming a cylindrical shape for the X-ray channel (Fig.~\ref{fig:shock}, left panel) and using the dependencies on the line of sight of the emission measure ($\propto L^{-1}$), pressure ($\propto L^{-1/2}$) and entropy ($\propto L^{1/3}$). For example, the ratio between the quantities measured at center and $r_5$ ($\equiv r_1$), \ie\ the distance of the outermost region, changes by $<16\%$ (or $<34\%$) for emission measure, $<8\%$ (or $<16\%$) for pressure and $<4\%$ (or $<10\%$) for entropy for cylinder radii $R > 2 r_5$ (or $> 1.5 r_5$).

\section{Discussion}

A1758 is an ideal object to study the merger processes between galaxy clusters and the impact of these events on their environment. Indeed, this system is composed of two main components, A1758N and A1758S, in different evolutionary stages (see Section~\ref{sec:a1758}). \\
\indent
The diffuse radio emission in A1758 follows the X-ray emission of the ICM (Fig.~\ref{fig:chandra_lofar_lowres}), suggesting a relation between the thermal and non-thermal components. The double cluster A1758N and A1758S represents the second system known to date to host two radio halos. The first discovered is the pair A399-A401 \citep{murgia10}, located at $z=0.07$. Both A1758 and A399-A401 show the presence of an X-ray channel between the cluster pair and some evidence for a lateral shock \citep{akamatsu17filament}. The bridge connecting A399 and A401 is more clear and indeed it has been also observed via the SZ effect by the \planck\ satellite \citep{planck13viii}; a hint of SZ signal is also found between A1758N and A1758S \citep{ami12}.

\subsection{The radio halos in the A1758 complex}\label{sec:relation}

\begin{figure}
 \centering
 \includegraphics[width=\hsize]{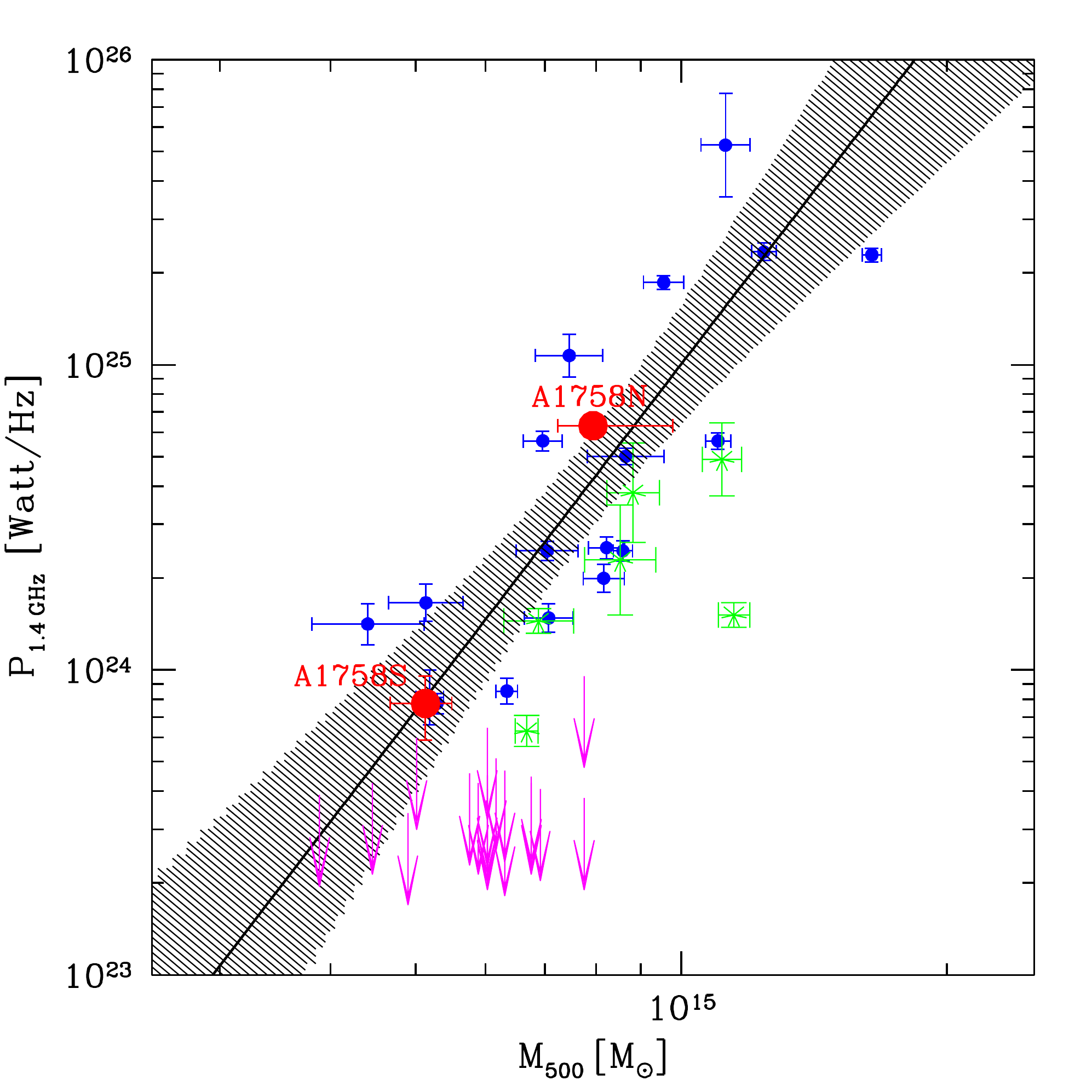}
 \caption{The $P_{1.4}-\mfive$ relation for giant radio halos. Different colors indicate: giant radio halos (\textit{blue}), ultra steep spectrum radio halos (\textit{green}), upper limits from \citet{venturi08} (\textit{magenta}) and the radio halos in A1758N and A1758S (\textit{red}). Errors on $P_{1.4}^{\rm N}$ are smaller than the point size. The best-fitting relation for giant radio halos and its 95\% confidence level are shown. Adapted from \citet{cassano13}.}
 \label{fig:relation}
\end{figure}

It is known that giant radio halos follow a relation between their radio power at 1.4~GHz and the mass of the hosting cluster \citep[\eg][]{cassano13}. We used the values reported in Tab.~\ref{tab:fluxes} to calculate the $k$-corrected 1.4~GHz radio power

\begin{equation}\label{eq:power}
 P_{1.4} = 4 \pi S_{\rm 1.4} D_L^2 (1+z)^{\alpha-1}
\end{equation}

\noindent
for the two halos in A1758N and A1758S, corresponding to $P_{1.4}^{\rm N} = (6.3\pm0.4) \times 10^{24}$ \whz\ and $P_{1.4}^{\rm S} = (7.7\pm1.8) \times 10^{23}$ \whz, respectively. \\
\indent
The $P_{\rm 1.4}-\mfive$ relation reported in \citet{cassano13} was obtained using the masses derived from the \planck\ satellite via SZ effect, which is known to be a robust indicator of the cluster mass \citep[\eg][]{motl05, nagai06}. However, the accuracy of the mass estimate for A1758N and A1758S with \planck\ is hindered by the difficulty of properly separating the two SZ components. Although the mass for A1758N has been estimated with different techniques, it is still uncertain and there is a large scatter in the values reported in the literature \citep[see Tab.~1 in][]{monteirooliveira17a1758}. The mass of A1758S is even more uncertain due to the lack of literature studies focused on this sub-cluster. In this respect, we adopted the $M-T$ relation reported in \citet{arnaud05} to estimate \mfive. We used the temperatures reported in \citet{david04}, and derived $M_{500}^{\rm N} = 8.0^{+1.8}_{-0.8}\times10^{14}$ \msun\ and $M_{500}^{\rm S} = (5.1\pm0.4)\times10^{14}$ \msun\ for A1758N and A1758S, respectively. We assumed these values being aware of the possible biases introduced in the scaling relation due to the ongoing mergers in A1758N and A1758S. However, we note that the masses estimated in such a way are within the values reported in the literature. \\
\indent
In Fig.~\ref{fig:relation} we compare our results with the $P_{\rm 1.4}-\mfive$ relation reported in \citet{cassano13}. The two radio halos in the A1758 complex lie very close to the best-fitting curve. Our results are in agreement with the fact that the most powerful radio halos are found in the most massive clusters.

\subsection{Merger scenario between A1758N and A1758S}\label{sec:merger_scenario}

\citet{david04} did not find any excess emission in the \xmm\ data in the region between A1758N and A1758S above that expected from a projection of the two systems. This suggested that the two components are not interacting because numerical simulations of merging clusters predict a surface brightness enhancement in the X-rays in the region of interaction \citep[\eg][]{roettiger97, ricker01, ritchie02}. However, our observations provide new insights on the merger scenario between A1758N and A1758S. \\
\indent
Thanks to the new and deep \chandra\ observations we were able to produce maps of the ICM thermodynamical quantities of all the A1758 complex (Fig.~\ref{fig:icm_maps}). They highlight the presence of high temperature and high entropy plasma in the region between the clusters, suggesting the existence of shock heated gas. This has been observed in a number of binary X-ray clusters in an early merging phase (\eg\ A98, \citealt{paternomahler14}; A115, \citealt{gutierrez05}; A141, \citealt{caglar18}; A399-A401, \citealt{akamatsu17filament}; A1750, \citealt{belsole04}; A3395, \citealt{lakhchaura11}; A3653, \citealt{caglar17}; 1E2216.0-0401 and 1E2215.7-0404, \citealt{akamatsu16}; CIZA J1358.9-4750, \citealt{kato15}) and it is in agreement with predictions by numerical simulations \citep[\eg][]{takizawa99, akahori10}. In contrast, the temperature enhancement is typically not observed when the separation of the pair exceeds their combined virial radii (\eg\ A2467 \citealt{wegner17}; A3528, \citealt{gastaldello03}; A3556-A3558, \citealt{mitsuishi12}; A3716, \citealt{andradesantos15}). Further indication of compressed gas in A1758 is given by the transversal profiles of Fig.~\ref{fig:shock} that also pinpoint a drop of the computed quantities outside the X-ray channel, toward the east direction. We speculated that this region traces a transversal shock. Reasons for this include the high Mach number inferred from the temperature jump ($\machkt = 3.0^{+1.4}_{-1.0}$) and its position resembling that of the ``equatorial'' shocks recently studied by \citet{ha18}. These shocks are the first to form during the merger phase and have high velocities and high Mach numbers since they propagate in very low density regions, contrary to those found in between the cluster pairs that are weaker due to the high temperature of the central medium \citep[\eg][]{belsole04, paternomahler14, kato15, akamatsu16, akamatsu17filament, caglar18}. \\
\indent
The $2\sigma$ level radio emission connecting A1758N and A1758S observed with \lofar\ (Fig.~\ref{fig:chandra_lofar_lowres}) needs further confirmation. If real, it could have been generated as a consequence of the encounter between the two clusters. This may indicate that part of the gravitational energy  is dissipated into non-thermal components during the early phase of the merger. We find intriguing its co-location with the possible transversal shock suggested by the temperature profile in Fig.~\ref{fig:shock} (see also Fig.~\ref{fig:low_snr_kt}). A shock could indeed power the radio emission similarly to the case of radio relics, whilst equatorial shocks are less energetic due to the lower density of the upstream gas\footnote{The energy dissipated by shocks is $\propto n_u V_{sh}^3$, where $n_u$ is the upstream density and $V_{sh}$ is the shock speed.} \citep{ha18}. \\
\indent
In conclusion, the results coming from our radio/X-ray analysis are consistent with a scenario where A1758N and A1758S are in a pre-merger phase, where the clusters are approaching, the gas between them is compressed and heated and the first shocks are launched. The application of a two-body dynamical model \citep[\eg][]{beers82} to test the gravitational binding of the clusters would be of great interest to probe the merging scenario; however, this is beyond the scope of this paper. Due to the overall complex dynamics of the merger (collision between clusters that are undergoing their own mergers), tailored numerical simulations would be useful to determine the impact velocities of the components in combination with multi-wavelength data (see \eg\ \citealt{molnar13} for the A1750 case).

\section{Conclusions}

We have presented new \lofar\ HBA observations of the double galaxy cluster A1758. In combination with archival \vla\ and \gmrt\ data, we have constrained the spectral properties of the diffuse radio emission in the ICM. We also analyzed a deep archival \chandra\ observation on the system. Here, we summarize our main results.

\begin{enumerate}
 \item The radio halo in A1758N is well known in the literature. \lofar\ has allowed us to recover diffuse radio emission from the ICM on a largest linear scale of $\sim2.2$ Mpc. The integrated spectral index computed from 144~MHz to 1.4~GHz is $\alpha = 1.2 \pm 0.1$. The radio power of this halo is $P_{1.4}^{\rm N} = (6.3\pm0.4) \times 10^{24}$ \whz. \\
 
 \item Using \lofar\ we have discovered a new, faint, radio halo in A1758S, which was not previously identified in studies at higher frequencies with the \gmrt\ and \vla\ observations. Our reanalysis of these datasets revealed its elusive nature and constrained its spectral index between 144~MHz and 1.4~GHz to $\alpha = 1.1 \pm 0.1$. The radio power of this halo is $P_{1.4}^{\rm S} = (7.7\pm1.8) \times 10^{23}$ \whz. \\
 
 \item Peripheral emission in the eastern outskirts of A1758S is also observed with \lofar, \gmrt\ and \vla. We tentatively classified this source as a radio relic ($\alpha = 1.3 \pm 0.1$). Although the relic origin is suggested by a number of observational properties (\eg\ morphology, location, linear extension), further observations are required to firmly determine its nature. \\
 
 \item The two radio halos in A1758N and A1758S lie within the 95\% confidence region of the best-fitting $P_{\rm 1.4}-\mfive$ relation reported by \citet{cassano13}. \\

 \item The maps of the ICM thermodynamical quantities computed from the deep \chandra\ observation indicate that the region between A1758N and A1758S is unrelaxed. In this respect, we suggested that the two sub-clusters are in a pre-merger phase. \\ 
 
 \item A possible bridge of radio emission connecting A1758N and A1758S is suggested by the low resolution \lofar\ image. The ICM temperature across this bridge shows a drop possibly indicating the presence of a compressed region or a transversal shock generated in the initial stage of the merger that could play a role in the formation of this diffuse emission. 
\end{enumerate}

\section*{Acknowledgments}

ABot thanks V.~Cuciti for helpful advises during the radio data analysis, and F.~Bedosti, T.~J.~Dijkema and D.~Rafferty for technical support on the LOFAR software installation. We thank F.~Gastaldello, C.~Haines and F.~Vazza for useful discussions, and the anonymous referee for suggestions that improved the manuscript. ABon acknowledges support from the ERC-Stg DRANOEL, no 714245. RJvW, DNH and HJAR acknowledge support from the ERC Advanced Investigator programme NewClusters 321271. RJvW acknowledges support from the VIDI research programme with project number 639.042.729, which is financed by the Netherlands Organisation for Scientific Research (NWO). FdG is supported by the VENI research programme with project number 1808, which is financed by the Netherlands Organisation for Scientific Research (NWO).  GJW gratefully acknowledges financial support from The Leverhulme Trust. This paper is based (in part) on data obtained with the International LOFAR Telescope (ILT) under project code LC2\_038. LOFAR \citep{vanhaarlem13} is the LOw Frequency ARray designed and constructed by ASTRON. It has observing, data processing, and data storage facilities in several countries, that are owned by various parties (each with their own funding sources), and that are collectively operated by the ILT foundation under a joint scientific policy. The ILT resources have benefitted from the following recent major funding sources: CNRS-INSU, Observatoire de Paris and Universit\'{e} d'Orl\'{e}ans, France; BMBF, MIWF-NRW, MPG, Germany; Science Foundation Ireland (SFI), Department of Business, Enterprise and Innovation (DBEI), Ireland; NWO, The Netherlands; The Science and Technology Facilities Council, UK. This work had made use of the LOFAR Solution Tool (LoSoTo), developed by F.~de Gasperin. We thank the staff of the GMRT that made these observations possible. GMRT is run by the National Centre for Radio Astrophysics of the Tata Institute of Fundamental Research. The NRAO is a facility of the National Science Foundation operated under cooperative agreement by Associated Universities, Inc. The scientific results reported in this article are based on observations made by the \chandra\ X-ray Observatory. This research made use of APLpy, an open-source plotting package for Python hosted at http://aplpy.github.com.

\bibliographystyle{mn2e}
\bibliography{library.bib}

\begin{thebibliography}{124}
\expandafter\ifx\csname natexlab\endcsname\relax\def\natexlab#1{#1}\fi

\bibitem[{Akahori \& Yoshikawa(2010)}]{akahori10}
Akahori T., Yoshikawa K., 2010, PASJ, 62, 335

\bibitem[{Akamatsu {et~al}\mbox{.}(2017)Akamatsu, Fujita, Akahori, Ishisaki,
  Hayashida, Hoshino, Mernier, Yoshikawa, Sato, \&
  Kaastra}]{akamatsu17filament}
Akamatsu H. {et~al.}, 2017, A\&A, 606, A1

\bibitem[{Akamatsu {et~al}\mbox{.}(2016)Akamatsu, Gu, Shimwell, Mernier, Mao,
  Urdampilleta, de~Plaa, R{\"{o}}ttgering, \& Kaastra}]{akamatsu16}
Akamatsu H. {et~al.}, 2016, A\&A, 593, L7

\bibitem[{Akamatsu \& Kawahara(2013)}]{akamatsu13systematic}
Akamatsu H., Kawahara H., 2013, PASJ, 65, 16

\bibitem[{{AMI Consortium: Rodr{\`{i}}guez-Gonz{\'{a}}lvez}
  {et~al}\mbox{.}(2012){AMI Consortium: Rodr{\`{i}}guez-Gonz{\'{a}}lvez},
  Shimwell, Davies, Feroz, Franzen, Grainge, Hobson, Hurley-Walker, Lasenby,
  Olamaie, Pooley, Saunders, Scaife, Schammel, Scott, Titterington, \&
  Waldram}]{ami12}
{AMI Consortium: Rodr{\`{i}}guez-Gonz{\'{a}}lvez} C. {et~al.}, 2012, MNRAS,
  425, 162

\bibitem[{Anders \& Grevesse(1989)}]{anders89}
Anders E., Grevesse N., 1989, Geochim. Cosmochim. Acta, 53, 197

\bibitem[{Andrade-Santos {et~al}\mbox{.}(2015)Andrade-Santos, Jones, Forman,
  Murray, Kraft, Vikhlinin, van Weeren, Nulsen, David, Dawson, Arnaud,
  Pointecouteau, Pratt, \& Melin}]{andradesantos15}
Andrade-Santos F. {et~al.}, 2015, ApJ, 803, 108

\bibitem[{Araya-Melo {et~al}\mbox{.}(2012)Araya-Melo, Arag{\'{o}}n-Calvo,
  Br{\"{u}}ggen, \& Hoeft}]{arayamelo12}
Araya-Melo P., Arag{\'{o}}n-Calvo M., Br{\"{u}}ggen M., Hoeft M., 2012, MNRAS,
  423, 2325

\bibitem[{Arnaud(1996)}]{arnaud96}
Arnaud K., 1996, in Astronomical Society of the Pacific Conference Series, Vol.
  101, Astron. Data Anal. Softw. Syst. V, Jacoby G., Barnes J., eds., p.~17

\bibitem[{Arnaud, Pointecouteau \& Pratt(2005)Arnaud, Pointecouteau, \&
  Pratt}]{arnaud05}
Arnaud M., Pointecouteau E., Pratt G., 2005, A\&A, 441, 893

\bibitem[{Bartalucci {et~al}\mbox{.}(2014)Bartalucci, Mazzotta, Bourdin, \&
  Vikhlinin}]{bartalucci14}
Bartalucci I., Mazzotta P., Bourdin H., Vikhlinin A., 2014, A\&A, 566, A25

\bibitem[{Beers, Geller \& Huchra(1982)Beers, Geller, \& Huchra}]{beers82}
Beers T., Geller M., Huchra J., 1982, ApJ, 257, 23

\bibitem[{Belsole {et~al}\mbox{.}(2004)Belsole, Pratt, Sauvageot, \&
  Bourdin}]{belsole04}
Belsole E., Pratt G., Sauvageot J.-L., Bourdin H., 2004, A\&A, 415, 821

\bibitem[{Bonafede {et~al}\mbox{.}(2014)Bonafede, Intema, Br{\"{u}}ggen,
  Girardi, Nonino, Kantharia, van Weeren, \&
  R{\"{o}}ttgering}]{bonafede14reacc}
Bonafede A., Intema H., Br{\"{u}}ggen M., Girardi M., Nonino M., Kantharia N.,
  van Weeren R., R{\"{o}}ttgering H., 2014, ApJ, 785, 1

\bibitem[{Boschin {et~al}\mbox{.}(2012)Boschin, Girardi, Barrena, \&
  Nonino}]{boschin12a1758}
Boschin W., Girardi M., Barrena R., Nonino M., 2012, A\&A, 540, A43

\bibitem[{Botteon, Gastaldello \& Brunetti(2018)Botteon, Gastaldello, \&
  Brunetti}]{botteon18edges}
Botteon A., Gastaldello F., Brunetti G., 2018, MNRAS, 476, 5591

\bibitem[{Botteon {et~al}\mbox{.}(2016{\natexlab{a}})Botteon, Gastaldello,
  Brunetti, \& Dallacasa}]{botteon16a115}
Botteon A., Gastaldello F., Brunetti G., Dallacasa D., 2016{\natexlab{a}},
  MNRAS, 460, L84

\bibitem[{Botteon {et~al}\mbox{.}(2016{\natexlab{b}})Botteon, Gastaldello,
  Brunetti, \& Kale}]{botteon16gordo}
Botteon A., Gastaldello F., Brunetti G., Kale R., 2016{\natexlab{b}}, MNRAS,
  463, 1534

\bibitem[{Brienza {et~al}\mbox{.}(2016)Brienza, Godfrey, Morganti, Vilchez,
  Maddox, Murgia, Orru, Shulevski, Best, Br{\"{u}}ggen, Harwood, Jamrozy,
  Jarvis, Mahony, McKean, \& R{\"{o}}ttgering}]{brienza16}
Brienza M. {et~al.}, 2016, A\&A, 585, A29

\bibitem[{Briggs(1995)}]{briggs95}
Briggs D., 1995, in Bulletin of the American Astronomical Society, Vol.~27,
  American Astronomical Society Meeting Abstracts, p. 1444

\bibitem[{Brunetti(2016)}]{brunetti16challenge}
Brunetti G., 2016, Plasma Phys. Control. Fusion, 58, 14011

\bibitem[{Brunetti \& Jones(2014)}]{brunetti14rev}
Brunetti G., Jones T., 2014, IJMPD, 23, 30007

\bibitem[{Brunetti \& Lazarian(2011)}]{brunetti11}
Brunetti G., Lazarian A., 2011, MNRAS, 410, 127

\bibitem[{Brunetti \& Lazarian(2016)}]{brunetti16stochastic}
Brunetti G., Lazarian A., 2016, MNRAS, 458, 2584

\bibitem[{Brunetti {et~al}\mbox{.}(2001)Brunetti, Setti, Feretti, \&
  Giovannini}]{brunetti01coma}
Brunetti G., Setti G., Feretti L., Giovannini G., 2001, MNRAS, 320, 365

\bibitem[{Brunetti {et~al}\mbox{.}(2007)Brunetti, Venturi, Dallacasa, Cassano,
  Dolag, Giacintucci, \& Setti}]{brunetti07cr}
Brunetti G., Venturi T., Dallacasa D., Cassano R., Dolag K., Giacintucci S.,
  Setti G., 2007, ApJ, 670, L5

\bibitem[{Brunetti, Zimmer \& Zandanel(2017)Brunetti, Zimmer, \&
  Zandanel}]{brunetti17}
Brunetti G., Zimmer S., Zandanel F., 2017, MNRAS, 472, 1506

\bibitem[{Buote(2001)}]{buote01}
Buote D., 2001, ApJ, 553, L15

\bibitem[{Caglar(2018)}]{caglar18}
Caglar T., 2018, MNRAS, 475, 2870

\bibitem[{Caglar \& Hudaverdi(2017)}]{caglar17}
Caglar T., Hudaverdi M., 2017, MNRAS, 472, 2633

\bibitem[{Cash(1979)}]{cash79}
Cash W., 1979, ApJ, 228, 939

\bibitem[{Cassano {et~al}\mbox{.}(2010{\natexlab{a}})Cassano, Brunetti,
  R{\"{o}}ttgering, \& Br{\"{u}}ggen}]{cassano10lofar}
Cassano R., Brunetti G., R{\"{o}}ttgering H., Br{\"{u}}ggen M.,
  2010{\natexlab{a}}, A\&A, 509, A68

\bibitem[{Cassano {et~al}\mbox{.}(2013)Cassano, Ettori, Brunetti, Giacintucci,
  Pratt, Venturi, Kale, Dolag, \& Markevitch}]{cassano13}
Cassano R. {et~al.}, 2013, ApJ, 777, 141

\bibitem[{Cassano {et~al}\mbox{.}(2010{\natexlab{b}})Cassano, Ettori,
  Giacintucci, Brunetti, Markevitch, Venturi, \& Gitti}]{cassano10connection}
Cassano R., Ettori S., Giacintucci S., Brunetti G., Markevitch M., Venturi T.,
  Gitti M., 2010{\natexlab{b}}, ApJ, 721, L82

\bibitem[{Chandra, Ray \& Bhatnagar(2004)Chandra, Ray, \&
  Bhatnagar}]{chandra04}
Chandra P., Ray A., Bhatnagar S., 2004, ApJ, 612, 974

\bibitem[{Condon {et~al}\mbox{.}(1998)Condon, Cotton, Greisen, Yin, Perley,
  Taylor, \& Broderick}]{condon98}
Condon J., Cotton W., Greisen E., Yin Q., Perley R., Taylor G., Broderick J.,
  1998, AJ, 115, 1693

\bibitem[{Cornwell, Golap \& Bhatnagar(2005)Cornwell, Golap, \&
  Bhatnagar}]{cornwell05}
Cornwell T., Golap K., Bhatnagar S., 2005, in Astronomical Society of the
  Pacific Conference Series, Vol. 347, Astronomical Data Analysis Software and
  Systems XIV, Shopbell P., Britton M., Ebert R., eds., p.~86

\bibitem[{Cuciti {et~al}\mbox{.}(2015)Cuciti, Cassano, Brunetti, Dallacasa,
  Kale, Ettori, \& Venturi}]{cuciti15}
Cuciti V., Cassano R., Brunetti G., Dallacasa D., Kale R., Ettori S., Venturi
  T., 2015, A\&A, 580, A97

\bibitem[{Dahle {et~al}\mbox{.}(2002)Dahle, Kaiser, Irgens, Lilje, \&
  Maddox}]{dahle02}
Dahle H., Kaiser N., Irgens R., Lilje P., Maddox S., 2002, ApJS, 139, 313

\bibitem[{David \& Kempner(2004)}]{david04}
David L., Kempner J., 2004, ApJ, 613, 831

\bibitem[{de~Gasperin {et~al}\mbox{.}(2017)de~Gasperin, Intema, Shimwell,
  Brunetti, Br{\"{u}}ggen, En{\ss}lin, van Weeren, Bonafede, \&
  R{\"{o}}ttgering}]{degasperin17gentle}
de~Gasperin F. {et~al.}, 2017, Science Adv., 3, e1701634

\bibitem[{Drabent(2017)}]{drabent17}
Drabent A., 2017, PhD thesis, Friedrich-Schiller-Universit{\"{a}}t Jena

\bibitem[{En{\ss}lin {et~al}\mbox{.}(1998)En{\ss}lin, Biermann, Klein, Kohle,
  Biermann, Klein, \& Kohle}]{ensslin98}
En{\ss}lin T., Biermann P., Klein U., Kohle S., Biermann P., Klein U., Kohle
  S., 1998, A\&A, 332, 395

\bibitem[{Feretti {et~al}\mbox{.}(2012)Feretti, Giovannini, Govoni, \&
  Murgia}]{feretti12rev}
Feretti L., Giovannini G., Govoni F., Murgia M., 2012, A\&A Rev., 20, 54

\bibitem[{Gastaldello {et~al}\mbox{.}(2003)Gastaldello, Ettori, Molendi,
  Bardelli, Venturi, \& Zucca}]{gastaldello03}
Gastaldello F., Ettori S., Molendi S., Bardelli S., Venturi T., Zucca E., 2003,
  A\&A, 411, 21

\bibitem[{Giovannini {et~al}\mbox{.}(2009)Giovannini, Bonafede, Feretti,
  Govoni, Murgia, Ferrari, \& Monti}]{giovannini09}
Giovannini G., Bonafede A., Feretti L., Govoni F., Murgia M., Ferrari F., Monti
  G., 2009, A\&A, 507, 1257

\bibitem[{Guo, Sironi \& Narayan(2014{\natexlab{a}})Guo, Sironi, \&
  Narayan}]{guo14a}
Guo X., Sironi L., Narayan R., 2014{\natexlab{a}}, ApJ, 794, 153

\bibitem[{Guo, Sironi \& Narayan(2014{\natexlab{b}})Guo, Sironi, \&
  Narayan}]{guo14b}
Guo X., Sironi L., Narayan R., 2014{\natexlab{b}}, ApJ, 797, 47

\bibitem[{Gutierrez \& Krawczynski(2005)}]{gutierrez05}
Gutierrez K., Krawczynski H., 2005, ApJ, 619, 161

\bibitem[{Ha, Ryu \& Kang(2018)Ha, Ryu, \& Kang}]{ha18}
Ha J.-H., Ryu D., Kang H., 2018, ApJ, 857, 26

\bibitem[{Haines {et~al}\mbox{.}(2017)Haines, Finoguenov, Smith, Babul, Egami,
  Mazzotta, Okabe, Pereira, Bianconi, McGee, Ziparo, Campusano, \&
  Loyola}]{haines17arx}
Haines C. {et~al.}, 2017, ArXiv e-prints

\bibitem[{Haines {et~al}\mbox{.}(2009)Haines, Smith, Egami, Okabe, Takada,
  Ellis, Moran, \& Umetsu}]{haines09}
Haines C., Smith G., Egami E., Okabe N., Takada M., Ellis R., Moran S., Umetsu
  K., 2009, MNRAS, 396, 1297

\bibitem[{Hardcastle {et~al}\mbox{.}(2016)Hardcastle, G{\"{u}}rkan, van Weeren,
  Williams, Best, de~Gasperin, Rafferty, Read, Sabater, Shimwell, Smith, Tasse,
  Bourne, Brienza, Br{\"{u}}ggen, Brunetti, Chy{\.{z}}y, Conway, Dunne, Eales,
  Maddox, Jarvis, Mahony, Morganti, Prandoni, R{\"{o}}ttgering, Valiante, \&
  White}]{hardcastle16}
Hardcastle M. {et~al.}, 2016, MNRAS, 462, 1910

\bibitem[{Hoang {et~al}\mbox{.}(2017)Hoang, Shimwell, Stroe, Akamatsu,
  Brunetti, Donnert, Intema, Mulcahy, R{\"{o}}ttgering, van Weeren, Bonafede,
  Br{\"{u}}ggen, Cassano, Chy{\.{z}}y, En{\ss}lin, Ferrari, de~Gasperin, Gu,
  Hoeft, Miley, Orr{\'{u}}, Pizzo, \& White}]{hoang17}
Hoang D. {et~al.}, 2017, MNRAS, 471, 1107

\bibitem[{Hoeft \& Br{\"{u}}ggen(2007)}]{hoeft07}
Hoeft M., Br{\"{u}}ggen M., 2007, MNRAS, 375, 77

\bibitem[{Intema {et~al}\mbox{.}(2017)Intema, Jagannathan, Mooley, \&
  Frail}]{intema17}
Intema H., Jagannathan P., Mooley K., Frail D., 2017, A\&A, 598, A78

\bibitem[{Intema {et~al}\mbox{.}(2009)Intema, van~der Tol, Cotton, Cohen, van
  Bemmel, \& R{\"{o}}ttgering}]{intema09}
Intema H., van~der Tol S., Cotton W., Cohen A., van Bemmel I., R{\"{o}}ttgering
  H., 2009, A\&A, 501, 1185

\bibitem[{Kalberla {et~al}\mbox{.}(2005)Kalberla, Burton, Hartmann, Arnal,
  Bajaja, Morras, \& P{\"{o}}ppel}]{kalberla05}
Kalberla P., Burton W., Hartmann D., Arnal E., Bajaja E., Morras R.,
  P{\"{o}}ppel W., 2005, A\&A, 440, 775

\bibitem[{Kang {et~al}\mbox{.}(2014)Kang, Petrosian, Ryu, \& Jones}]{kang14}
Kang H., Petrosian V., Ryu D., Jones T., 2014, ApJ, 788, 142

\bibitem[{Kang \& Ryu(2011)}]{kang11}
Kang H., Ryu D., 2011, ApJ, 734, 18

\bibitem[{Kang \& Ryu(2016)}]{kang16reacc}
Kang H., Ryu D., 2016, ApJ, 823, 13

\bibitem[{Kang, Ryu \& Jones(2012)Kang, Ryu, \& Jones}]{kang12}
Kang H., Ryu D., Jones T., 2012, ApJ, 756, 97

\bibitem[{Kato {et~al}\mbox{.}(2015)Kato, Nakazawa, Gu, Akahori, Takizawa,
  Fujita, \& Makishima}]{kato15}
Kato Y., Nakazawa K., Gu L., Akahori T., Takizawa M., Fujita Y., Makishima K.,
  2015, PASJ, 67, 71

\bibitem[{Kempner \& Sarazin(2001)}]{kempner01}
Kempner J., Sarazin C., 2001, ApJ, 548, 639

\bibitem[{Keshet, Waxman \& Loeb(2004)Keshet, Waxman, \&
  Loeb}]{keshet04imprint}
Keshet U., Waxman E., Loeb A., 2004, ApJ, 617, 281

\bibitem[{Lakhchaura {et~al}\mbox{.}(2011)Lakhchaura, Singh, Saikia, \&
  Hunstead}]{lakhchaura11}
Lakhchaura K., Singh K., Saikia D., Hunstead R., 2011, ApJ, 743, 78

\bibitem[{Landau \& Lifshitz(1959)}]{landau59}
Landau L., Lifshitz E., 1959, {Fluid mechanics}

\bibitem[{Lane {et~al}\mbox{.}(2014)Lane, Cotton, van Velzen, Clarke, Kassim,
  Helmboldt, Lazio, \& Cohen}]{lane14}
Lane W., Cotton W., van Velzen S., Clarke T., Kassim N., Helmboldt J., Lazio
  T., Cohen A., 2014, MNRAS, 440, 327

\bibitem[{Macario {et~al}\mbox{.}(2011)Macario, Markevitch, Giacintucci,
  Brunetti, Venturi, \& Murray}]{macario11}
Macario G., Markevitch M., Giacintucci S., Brunetti G., Venturi T., Murray S.,
  2011, ApJ, 728, 82

\bibitem[{Machado {et~al}\mbox{.}(2015)Machado, Monteiro-Oliveira, {Lima Neto},
  \& Cypriano}]{machado15a1758}
Machado R., Monteiro-Oliveira R., {Lima Neto} G., Cypriano E., 2015, MNRAS,
  451, 3309

\bibitem[{Markevitch \& Vikhlinin(2007)}]{markevitch07rev}
Markevitch M., Vikhlinin A., 2007, Phys. Rep., 443, 1

\bibitem[{Martino {et~al}\mbox{.}(2014)Martino, Mazzotta, Bourdin, Smith,
  Bartalucci, Marrone, Finoguenov, \& Okabe}]{martino14}
Martino R., Mazzotta P., Bourdin H., Smith G., Bartalucci I., Marrone D.,
  Finoguenov A., Okabe N., 2014, MNRAS, 443, 2342

\bibitem[{Mazzotta {et~al}\mbox{.}(2004)Mazzotta, Rasia, Moscardini, \&
  Tormen}]{mazzotta04}
Mazzotta P., Rasia E., Moscardini L., Tormen G., 2004, MNRAS, 354, 10

\bibitem[{McMullin {et~al}\mbox{.}(2007)McMullin, Waters, Schiebel, Young, \&
  Golap}]{mcmullin07}
McMullin J., Waters B., Schiebel D., Young W., Golap K., 2007, in Astronomical
  Society of the Pacific Conference Series, Vol. 376, Astronomical Data
  Analysis Software and Systems XVI, Shaw R., Hill F., Bell D., eds., p. 127

\bibitem[{Mitsuishi {et~al}\mbox{.}(2012)Mitsuishi, Gupta, Yamasaki, Takei,
  Ohashi, Sato, Galeazzi, Henry, \& Kelley}]{mitsuishi12}
Mitsuishi I. {et~al.}, 2012, PASJ, 64, 18

\bibitem[{Mohan \& Rafferty(2015)}]{mohan15}
Mohan N., Rafferty D., 2015, {PyBDSF: Python Blob Detection and Source Finder}.
  Astrophysics Source Code Library

\bibitem[{Molnar {et~al}\mbox{.}(2013)Molnar, Chiu, Broadhurst, \&
  Stadel}]{molnar13}
Molnar S., Chiu I.-N., Broadhurst T., Stadel J., 2013, ApJ, 779, 63

\bibitem[{Monteiro-Oliveira {et~al}\mbox{.}(2017)Monteiro-Oliveira, Cypriano,
  Machado, {Lima Neto}, Ribeiro, Sodr{\'{e}}, \&
  Dupke}]{monteirooliveira17a1758}
Monteiro-Oliveira R., Cypriano E., Machado R., {Lima Neto} G., Ribeiro A.,
  Sodr{\'{e}} L., Dupke R., 2017, MNRAS, 466, 2614

\bibitem[{Motl {et~al}\mbox{.}(2005)Motl, Hallman, Burns, \& Norman}]{motl05}
Motl P., Hallman E., Burns J., Norman M., 2005, ApJ, 623, L63

\bibitem[{Murgia {et~al}\mbox{.}(2010)Murgia, Govoni, Feretti, \&
  Giovannini}]{murgia10}
Murgia M., Govoni F., Feretti L., Giovannini G., 2010, A\&A, 509, A86

\bibitem[{Nagai(2006)}]{nagai06}
Nagai D., 2006, ApJ, 650, 538

\bibitem[{Nuza {et~al}\mbox{.}(2012)Nuza, Hoeft, van Weeren, Gottl{\"{o}}ber,
  \& Yepes}]{nuza12}
Nuza S., Hoeft M., van Weeren R., Gottl{\"{o}}ber S., Yepes G., 2012, MNRAS,
  420, 2006

\bibitem[{O'Dea \& Owen(1985)}]{odea85}
O'Dea C., Owen F., 1985, AJ, 90, 927

\bibitem[{Offringa {et~al}\mbox{.}(2014)Offringa, McKinley, Hurley-Walker,
  Briggs, Wayth, Kaplan, Bell, Feng, Neben, Hughes, Rhee, Murphy, Bhat,
  Bernardi, Bowman, Cappallo, Corey, Deshpande, Emrich, Ewall-Wice, Gaensler,
  Goeke, Greenhill, Hazelton, Hindson, Johnston-Hollitt, Jacobs, Kasper,
  Kratzenberg, Lenc, Lonsdale, Lynch, McWhirter, Mitchell, Morales, Morgan,
  Kudryavtseva, Oberoi, Ord, Pindor, Procopio, Prabu, Riding, Roshi, Shankar,
  Srivani, Subrahmanyan, Tingay, Waterson, Webster, Whitney, Williams, \&
  Williams}]{offringa14}
Offringa A. {et~al.}, 2014, MNRAS, 444, 606

\bibitem[{Okabe \& Umetsu(2008)}]{okabe08}
Okabe N., Umetsu K., 2008, PASJ, 60, 345

\bibitem[{Paterno-Mahler {et~al}\mbox{.}(2014)Paterno-Mahler, Randall, Bulbul,
  Andrade-Santos, Blanton, Jones, Murray, \& Johnson}]{paternomahler14}
Paterno-Mahler R., Randall S., Bulbul E., Andrade-Santos F., Blanton E., Jones
  C., Murray S., Johnson R., 2014, ApJ, 791, 104

\bibitem[{Perley \& Butler(2013)}]{perley13}
Perley R., Butler B., 2013, ApJS, 204, 19

\bibitem[{Petrosian(2001)}]{petrosian01}
Petrosian V., 2001, ApJ, 557, 560

\bibitem[{Pinzke, Oh \& Pfrommer(2013)Pinzke, Oh, \& Pfrommer}]{pinzke13}
Pinzke A., Oh S., Pfrommer C., 2013, MNRAS, 435, 1061

\bibitem[{Pinzke, Oh \& Pfrommer(2017)Pinzke, Oh, \& Pfrommer}]{pinzke17}
Pinzke A., Oh S., Pfrommer C., 2017, MNRAS, 465, 4800

\bibitem[{{Planck Collaboration VIII}(2013)}]{planck13viii}
{Planck Collaboration VIII}, 2013, A\&A, 550, A134

\bibitem[{Ragozzine {et~al}\mbox{.}(2012)Ragozzine, Clowe, Markevitch,
  Gonzalez, \& Brada{\v{c}}}]{ragozzine12}
Ragozzine B., Clowe D., Markevitch M., Gonzalez A., Brada{\v{c}} M., 2012, ApJ,
  744, 94

\bibitem[{Rau \& Cornwell(2011)}]{rau11}
Rau U., Cornwell T., 2011, A\&A, 532, A71

\bibitem[{Rengelink {et~al}\mbox{.}(1997)Rengelink, Tang, de~Bruyn, Miley,
  Bremer, Roettgering, \& Bremer}]{rengelink97}
Rengelink R., Tang Y., de~Bruyn A., Miley G., Bremer M., Roettgering H., Bremer
  M., 1997, A\&AS, 124

\bibitem[{Ricker \& Sarazin(2001)}]{ricker01}
Ricker P., Sarazin C., 2001, ApJ, 561, 621

\bibitem[{Ritchie \& Thomas(2002)}]{ritchie02}
Ritchie B., Thomas P., 2002, MNRAS, 329, 675

\bibitem[{Rizza {et~al}\mbox{.}(1998)Rizza, Burns, Ledlow, Owen, Voges, \&
  Bliton}]{rizza98}
Rizza E., Burns J., Ledlow M., Owen F., Voges W., Bliton M., 1998, MNRAS, 301,
  328

\bibitem[{Roettiger, Burns \& Stone(1999)Roettiger, Burns, \&
  Stone}]{roettiger99a3667}
Roettiger K., Burns J., Stone J., 1999, ApJ, 518, 603

\bibitem[{Roettiger, Loken \& Burns(1997)Roettiger, Loken, \&
  Burns}]{roettiger97}
Roettiger K., Loken C., Burns J., 1997, A\&AS, 109, 307

\bibitem[{R{\"{o}}ttgering {et~al}\mbox{.}(2011)R{\"{o}}ttgering, Afonso,
  Barthel, Batejat, Best, Bonafede, Br{\"{u}}ggen, Brunetti, Chy{\.{z}}y,
  Conway, Gasperin, Ferrari, Haverkorn, Heald, Hoeft, Jackson, Jarvis, Ker,
  Lehnert, Macario, McKean, Miley, Morganti, Oosterloo, Orr{\`{u}}, Pizzo,
  Rafferty, Shulevski, Tasse, Bemmel, van~der Tol, van Weeren, Verheijen,
  White, \& Wise}]{rottgering11}
R{\"{o}}ttgering H. {et~al.}, 2011, J. Astrophys. Astron., 32, 557

\bibitem[{R{\"{o}}ttgering {et~al}\mbox{.}(2006)R{\"{o}}ttgering, Braun,
  Barthel, van Haarlem, Miley, Morganti, Snellen, Falcke, de~Bruyn, Stappers,
  Boland, Butcher, de~Geus, Koopmans, Fender, Kuijpers, Schilizzi, Vogt,
  Wijers, Wise, Brouw, Hamaker, Noordam, Oosterloo, Bahren, Brentjens,
  Wijnholds, Bregman, van Cappellen, Gunst, Kant, Reitsma, van~der Schaaf, \&
  de~Vos}]{rottgering06}
R{\"{o}}ttgering H. {et~al.}, 2006, ArXiv e-prints

\bibitem[{Sanders(2006)}]{sanders06contbin}
Sanders J., 2006, MNRAS, 371, 829

\bibitem[{Savini {et~al}\mbox{.}(2018)Savini, Bonafede, Br{\"{u}}ggen, Wilber,
  Harwood, Murgia, Shimwell, Rafferty, Shulevski, Brienza, Hardcastle,
  Morganti, R{\"{o}}ttgering, Clarke, de~Gasperin, van Weeren, Best, Botteon,
  Brunetti, \& Cassano}]{savini18}
Savini F. {et~al.}, 2018, MNRAS, 474, 5023

\bibitem[{Scaife \& Heald(2012)}]{scaife12}
Scaife A., Heald G., 2012, MNRAS, 423, L30

\bibitem[{Shimwell {et~al}\mbox{.}(2016)Shimwell, Luckin, Br{\"{u}}ggen,
  Brunetti, Intema, Owers, R{\"{o}}ttgering, Stroe, van Weeren, Williams,
  Cassano, de~Gasperin, Heald, Hoang, Hardcastle, Sridhar, Sabater, Best,
  Bonafede, Chy{\.{z}}y, En{\ss}lin, Ferrari, Haverkorn, Hoeft, Horellou,
  McKean, Morabito, Orr{\`{u}}, Pizzo, Retana-Montenegro, \&
  White}]{shimwell16}
Shimwell T. {et~al.}, 2016, MNRAS, 459, 277

\bibitem[{Shimwell {et~al}\mbox{.}(2015)Shimwell, Markevitch, Brown, Feretti,
  Gaensler, Johnston-Hollitt, Lage, \& Srinivasan}]{shimwell15}
Shimwell T., Markevitch M., Brown S., Feretti L., Gaensler B., Johnston-Hollitt
  M., Lage C., Srinivasan R., 2015, MNRAS, 449, 1486

\bibitem[{Shimwell {et~al}\mbox{.}(2017)Shimwell, R{\"{o}}ttgering, Best,
  Williams, Dijkema, de~Gasperin, Hardcastle, Heald, Hoang, Horneffer, Intema,
  Mahony, Mandal, Mechev, Morabito, Oonk, Rafferty, Retana-Montenegro, Sabater,
  Tasse, van Weeren, Br{\"{u}}ggen, Brunetti, Chy{\.{z}}y, Conway, Haverkorn,
  Jackson, Jarvis, McKean, Miley, Morganti, White, Wise, van Bemmel, Beck,
  Brienza, Bonafede, {Calistro Rivera}, Cassano, Clarke, Cseh, Deller, Drabent,
  van Driel, Engels, Falcke, Ferrari, Fr{\"{o}}hlich, Garrett, Harwood, Heesen,
  Hoeft, Horellou, Israel, Kapi{\'{n}}ska, Kunert-Bajraszewska, McKay, Mohan,
  Orr{\'{u}}, Pizzo, Prandoni, Schwarz, Shulevski, Sipior, Smith, Sridhar,
  Steinmetz, Stroe, Varenius, van~der Werf, Zensus, \& Zwart}]{shimwell17}
Shimwell T. {et~al.}, 2017, A\&A, 598, A104

\bibitem[{Sirothia(2009)}]{sirothia09}
Sirothia S., 2009, MNRAS, 398, 853

\bibitem[{Takizawa(1999)}]{takizawa99}
Takizawa M., 1999, ApJ, 520, 514

\bibitem[{Tasse {et~al}\mbox{.}(2013)Tasse, van~der Tol, van Zwieten, van
  Diepen, \& Bhatnagar}]{tasse13}
Tasse C., van~der Tol S., van Zwieten J., van Diepen G., Bhatnagar S., 2013,
  A\&A, 553, A105

\bibitem[{van Haarlem {et~al}\mbox{.}(2013)van Haarlem, Wise, Gunst, Heald,
  McKean, Hessels, de~Bruyn, Nijboer, Swinbank, Fallows, Brentjens, Nelles,
  Beck, Falcke, Fender, H{\"{o}}randel, Koopmans, Mann, Miley,
  R{\"{o}}ttgering, Stappers, Wijers, Zaroubi, van~den Akker, Alexov, Anderson,
  Anderson, van Ardenne, Arts, Asgekar, Avruch, Batejat, B{\"{a}}hren, Bell,
  Bell, van Bemmel, Bennema, Bentum, Bernardi, Best, B{\^{i}}rzan, Bonafede,
  Boonstra, Braun, Bregman, Breitling, van~de Brink, Broderick, Broekema,
  Brouw, Br{\"{u}}ggen, Butcher, van Cappellen, Ciardi, Coenen, Conway, Coolen,
  Corstanje, Damstra, Davies, Deller, Dettmar, van Diepen, Dijkstra, Donker,
  Doorduin, Dromer, Drost, van Duin, Eisl{\"{o}}ffel, van Enst, Ferrari,
  Frieswijk, Gankema, Garrett, de~Gasperin, Gerbers, de~Geus, Grie{\ss}meier,
  Grit, Gruppen, Hamaker, Hassall, Hoeft, Holties, Horneffer, van~der Horst,
  van Houwelingen, Huijgen, Iacobelli, Intema, Jackson, Jelic, de~Jong, Juette,
  Kant, Karastergiou, Koers, Kollen, Kondratiev, Kooistra, Koopman, Koster,
  Kuniyoshi, Kramer, Kuper, Lambropoulos, Law, van Leeuwen, Lemaitre, Loose,
  Maat, Macario, Markoff, Masters, McFadden, McKay-Bukowski, Meijering,
  Meulman, Mevius, Middelberg, Millenaar, Miller-Jones, Mohan, Mol, Morawietz,
  Morganti, Mulcahy, Mulder, Munk, Nieuwenhuis, van Nieuwpoort, Noordam,
  Norden, Noutsos, Offringa, Olofsson, Omar, Orr{\'{u}}, Overeem, Paas,
  Pandey-Pommier, Pandey, Pizzo, Polatidis, Rafferty, Rawlings, Reich,
  de~Reijer, Reitsma, Renting, Riemers, Rol, Romein, Roosjen, Ruiter, Scaife,
  van~der Schaaf, Scheers, Schellart, Schoenmakers, Schoonderbeek, Serylak,
  Shulevski, Sluman, Smirnov, Sobey, Spreeuw, Steinmetz, Sterks, Stiepel,
  Stuurwold, Tagger, Tang, Tasse, Thomas, Thoudam, Toribio, van~der Tol, Usov,
  van Veelen, van~der Veen, ter Veen, Verbiest, Vermeulen, Vermaas, Vocks,
  Vogt, de~Vos, van~der Wal, van Weeren, Weggemans, Weltevrede, White,
  Wijnholds, Wilhelmsson, Wucknitz, Yatawatta, Zarka, Zensus, \& van
  Zwieten}]{vanhaarlem13}
van Haarlem M. {et~al.}, 2013, A\&A, 556, A2

\bibitem[{van Weeren {et~al}\mbox{.}(2017)van Weeren, Andrade-Santos, Dawson,
  Golovich, Lal, Kang, Ryu, Br{\"{u}}ggen, Ogrean, Forman, Jones, Placco,
  Santucci, Wittman, Jee, Kraft, Sobral, Stroe, \& Fogarty}]{vanweeren17a3411}
van Weeren R. {et~al.}, 2017, Nature Astron., 1, 5

\bibitem[{van Weeren {et~al}\mbox{.}(2016{\natexlab{a}})van Weeren, Brunetti,
  Br{\"{u}}ggen, Andrade-Santos, Ogrean, Williams, R{\"{o}}ttgering, Dawson,
  Forman, de~Gasperin, Hardcastle, Jones, Miley, Rafferty, Rudnick, Sabater,
  Sarazin, Shimwell, Bonafede, Best, B{\^{i}}rzan, Cassano, Chy{\.{z}}y,
  Croston, Dijkema, En{\ss}lin, Ferrari, Heald, Hoeft, Horellou, Jarvis, Kraft,
  Mevius, Intema, Murray, Orr{\'{u}}, Pizzo, Sridhar, Simionescu, Stroe,
  van~der Tol, \& White}]{vanweeren16toothbrush}
van Weeren R. {et~al.}, 2016{\natexlab{a}}, ApJ, 818, 204

\bibitem[{van Weeren {et~al}\mbox{.}(2016{\natexlab{b}})van Weeren, Williams,
  Hardcastle, Shimwell, Rafferty, Sabater, Heald, Sridhar, Dijkema, Brunetti,
  Br{\"{u}}ggen, Andrade-Santos, Ogrean, R{\"{o}}ttgering, Dawson, Forman,
  de~Gasperin, Jones, Miley, Rudnick, Sarazin, Bonafede, Best, B{\^{i}}rzan,
  Cassano, Chy{\.{z}}y, Croston, En{\ss}lin, Ferrari, Hoeft, Horellou, Jarvis,
  Kraft, Mevius, Intema, Murray, Orr{\'{u}}, Pizzo, Simionescu, Stroe, van~der
  Tol, \& White}]{vanweeren16calibration}
van Weeren R. {et~al.}, 2016{\natexlab{b}}, ApJS, 223, 2

\bibitem[{Vazza \& Br{\"{u}}ggen(2014)}]{vazza14challenge}
Vazza F., Br{\"{u}}ggen M., 2014, MNRAS, 437, 2291

\bibitem[{Vazza {et~al}\mbox{.}(2016)Vazza, Br{\"{u}}ggen, Wittor, Gheller,
  Eckert, \& Stubbe}]{vazza16}
Vazza F., Br{\"{u}}ggen M., Wittor D., Gheller C., Eckert D., Stubbe M., 2016,
  MNRAS, 459, 70

\bibitem[{Vazza {et~al}\mbox{.}(2015{\natexlab{a}})Vazza, Eckert,
  Br{\"{u}}ggen, \& Huber}]{vazza15efficiency}
Vazza F., Eckert D., Br{\"{u}}ggen M., Huber B., 2015{\natexlab{a}}, MNRAS,
  451, 2198

\bibitem[{Vazza {et~al}\mbox{.}(2015{\natexlab{b}})Vazza, Ferrari,
  Br{\"{u}}ggen, Bonafede, Gheller, \& Wang}]{vazza15forecasts}
Vazza F., Ferrari C., Br{\"{u}}ggen M., Bonafede A., Gheller C., Wang P.,
  2015{\natexlab{b}}, A\&A, 580, A119

\bibitem[{Venturi {et~al}\mbox{.}(2008)Venturi, Giacintucci, Dallacasa,
  Cassano, Brunetti, Bardelli, \& Setti}]{venturi08}
Venturi T., Giacintucci S., Dallacasa D., Cassano R., Brunetti G., Bardelli S.,
  Setti G., 2008, A\&A, 484, 327

\bibitem[{Venturi {et~al}\mbox{.}(2013)Venturi, Giacintucci, Dallacasa,
  Cassano, Brunetti, Macario, \& Athreya}]{venturi13}
Venturi T., Giacintucci S., Dallacasa D., Cassano R., Brunetti G., Macario G.,
  Athreya R., 2013, A\&A, 551, A24

\bibitem[{Wegner {et~al}\mbox{.}(2017)Wegner, Umetsu, Molnar, Nonino,
  Medezinski, Andrade-Santos, Bogdan, Lovisari, Forman, \& Jones}]{wegner17}
Wegner G. {et~al.}, 2017, ApJ, 844, 67

\bibitem[{Wilber {et~al}\mbox{.}(2018)Wilber, Br{\"{u}}ggen, Bonafede, Savini,
  Shimwell, van Weeren, Rafferty, Mechev, Intema, Andrade-Santos, Clarke,
  Mahony, Morganti, Prandoni, Brunetti, R{\"{o}}ttgering, Mandal, de~Gasperin,
  \& Hoeft}]{wilber18a1132}
Wilber A. {et~al.}, 2018, MNRAS, 473, 3536

\bibitem[{Williams {et~al}\mbox{.}(2016)Williams, van Weeren, R{\"{o}}ttgering,
  Best, Dijkema, de~Gasperin, Hardcastle, Heald, Prandoni, Sabater, Shimwell,
  Tasse, van Bemmel, Br{\"{u}}ggen, Brunetti, Conway, En{\ss}lin, Engels,
  Falcke, Ferrari, Haverkorn, Jackson, Jarvis, Kapi{\'{n}}ska, Mahony, Miley,
  Morabito, Morganti, Orr{\'{u}}, Retana-Montenegro, Sridhar, Toribio, White,
  Wise, \& Zwart}]{williams16}
Williams W. {et~al.}, 2016, MNRAS, 460, 2385

\bibitem[{Wittor, Vazza \& Br{\"{u}}ggen(2017)Wittor, Vazza, \&
  Br{\"{u}}ggen}]{wittor17relics}
Wittor D., Vazza F., Br{\"{u}}ggen M., 2017, MNRAS, 464, 4448

\end{thebibliography}

\appendix

\section{Temperature map}\label{app:errormap}

\begin{figure}
 \centering
 \includegraphics[width=.8\hsize]{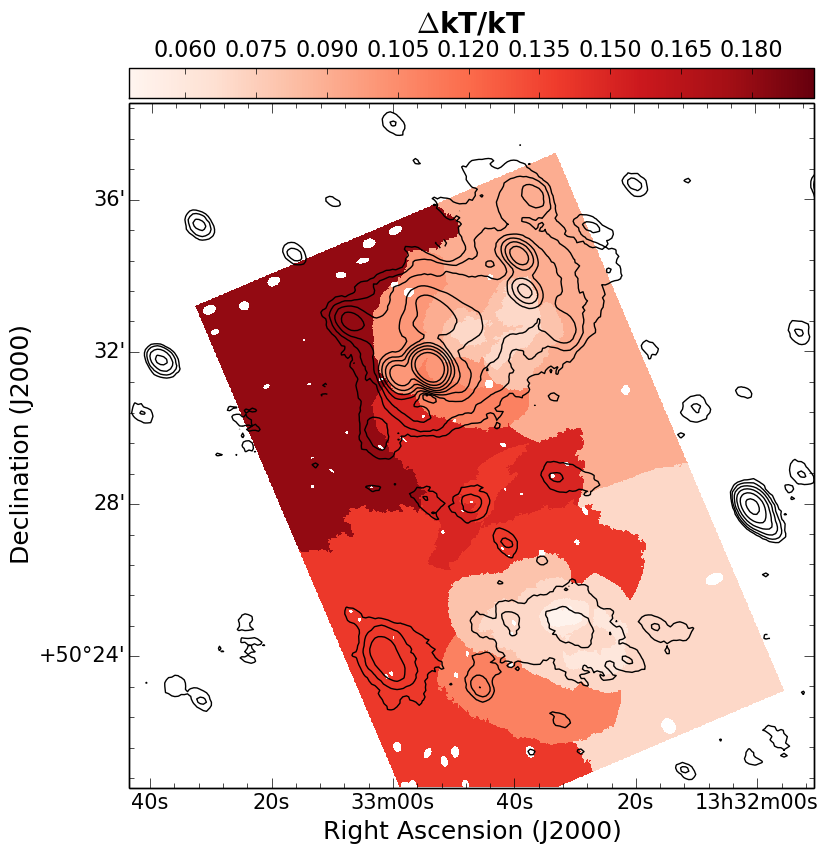}
 \caption{Temperature error map of A1758 (\cf\ with Fig~\ref{fig:icm_maps}).}
 \label{fig:kt_error}
\end{figure}

\begin{figure*}
 \centering
 \includegraphics[width=.4\textwidth]{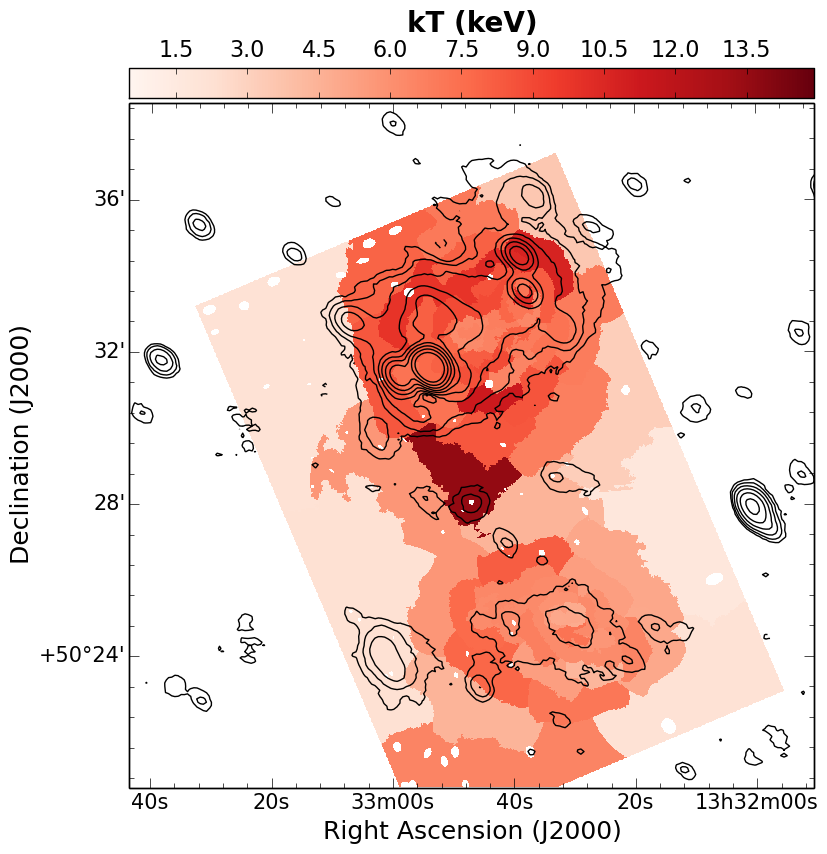}
 \includegraphics[width=.4\textwidth]{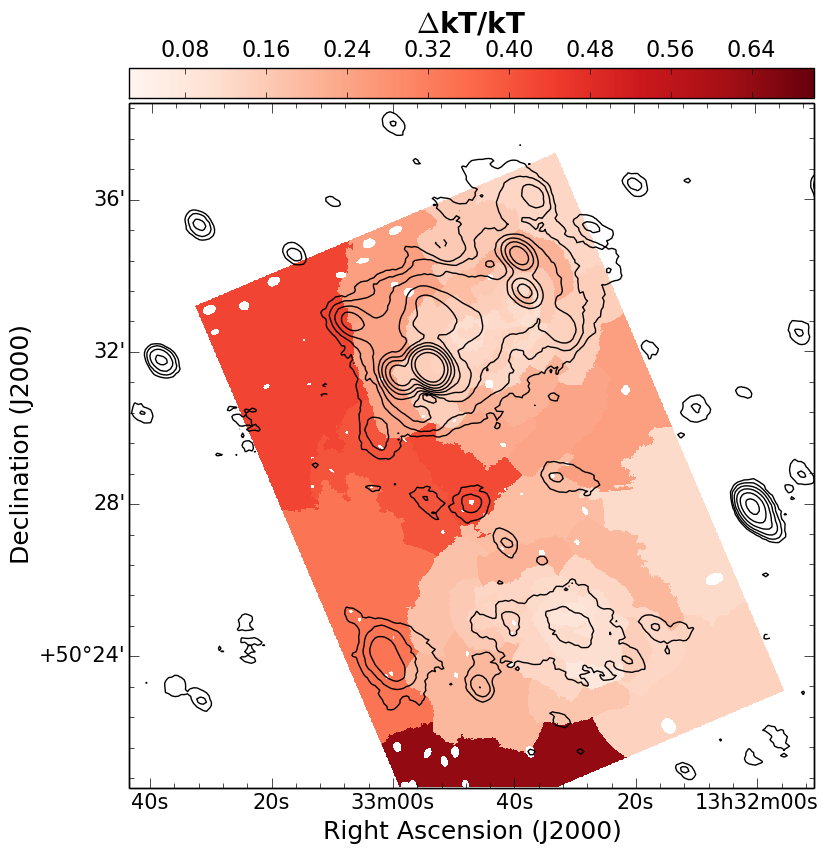}
 \caption{Low S/N temperature map (\textit{left}) and relative error map (\textit{right}) of A1758.}
 \label{fig:low_snr_kt}
\end{figure*}

The fractional errors on the temperature map of Fig~\ref{fig:icm_maps} are reported in Fig.~\ref{fig:kt_error}. Projected pressure and entropy have similar fractional errors due to their linear dependence on the temperature and to the general small error on the emission measure. \\
\indent
The high value of temperature in the putative post-shock region of Fig.~\ref{fig:shock} can not be identified in the temperature map of Fig.~\ref{fig:icm_maps} (left panel) likely due to the fact that the spectral extracting sectors are large, hence they might contain a mix of plasma with different temperatures. In Fig.~\ref{fig:low_snr_kt} we show that if the required S/N per region is reduced to 30, the \contbin\ algorithm is able to draw a smaller sector similar in size and position to region 2 in Fig.~\ref{fig:shock} where the spectral fit provides again $kT \sim 15$ \kev, canceling the apparent tension between the two results.

\section{X-ray channel spectra}\label{app:spectra}

\begin{figure*}
 \centering
 \includegraphics[width=.3\hsize,angle=-90]{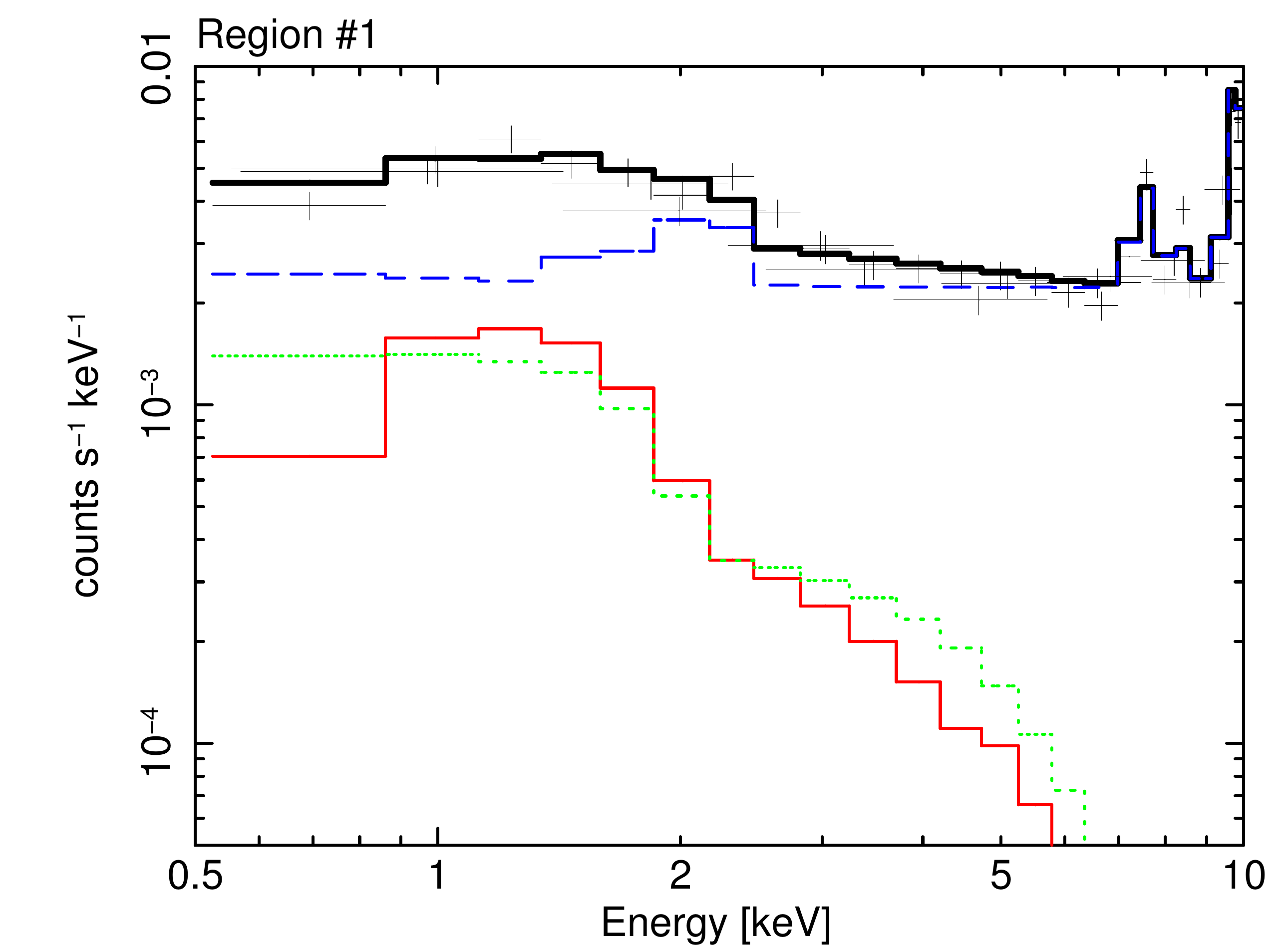}
 \includegraphics[width=.3\hsize,angle=-90]{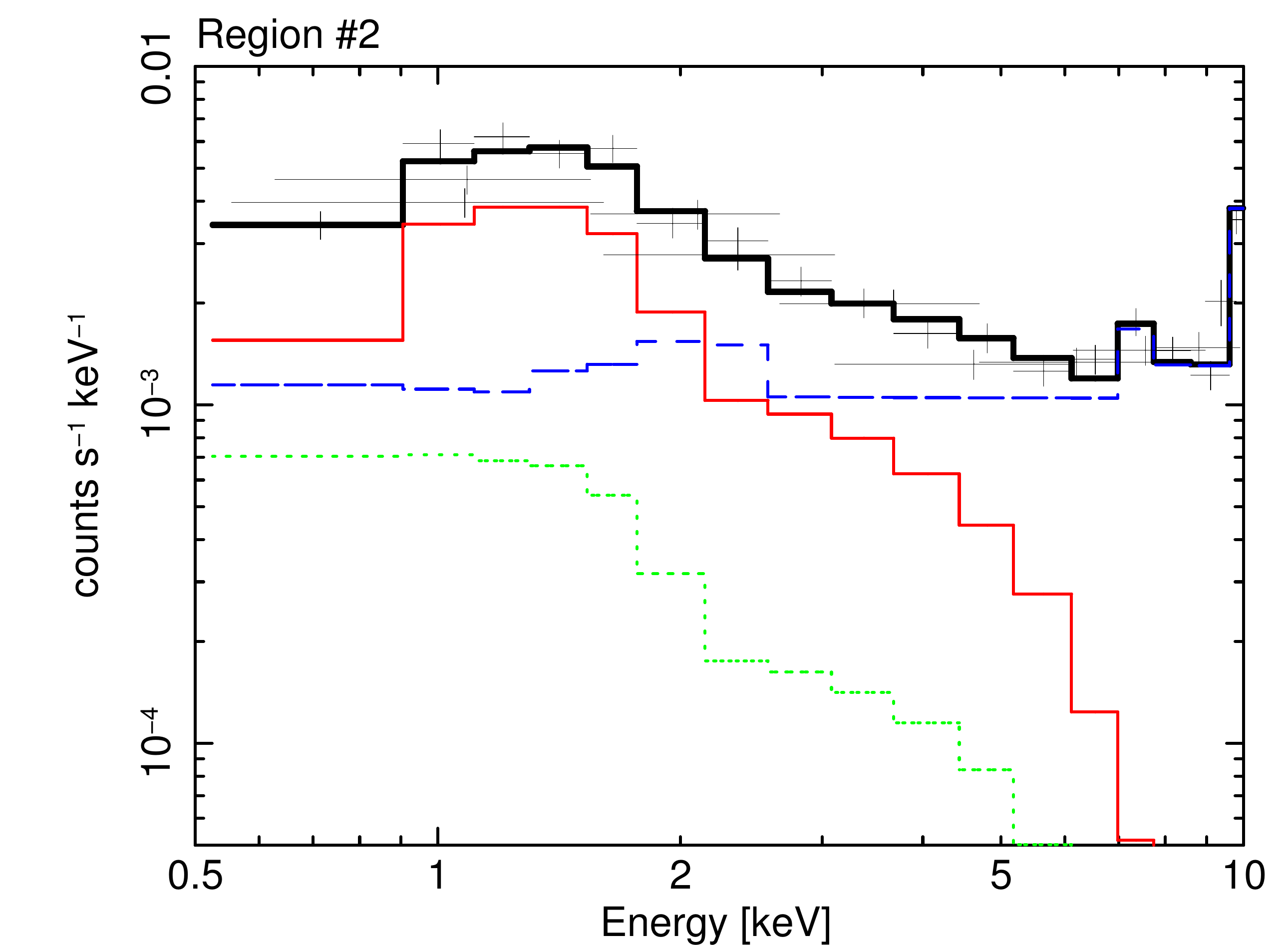}
 \includegraphics[width=.3\hsize,angle=-90]{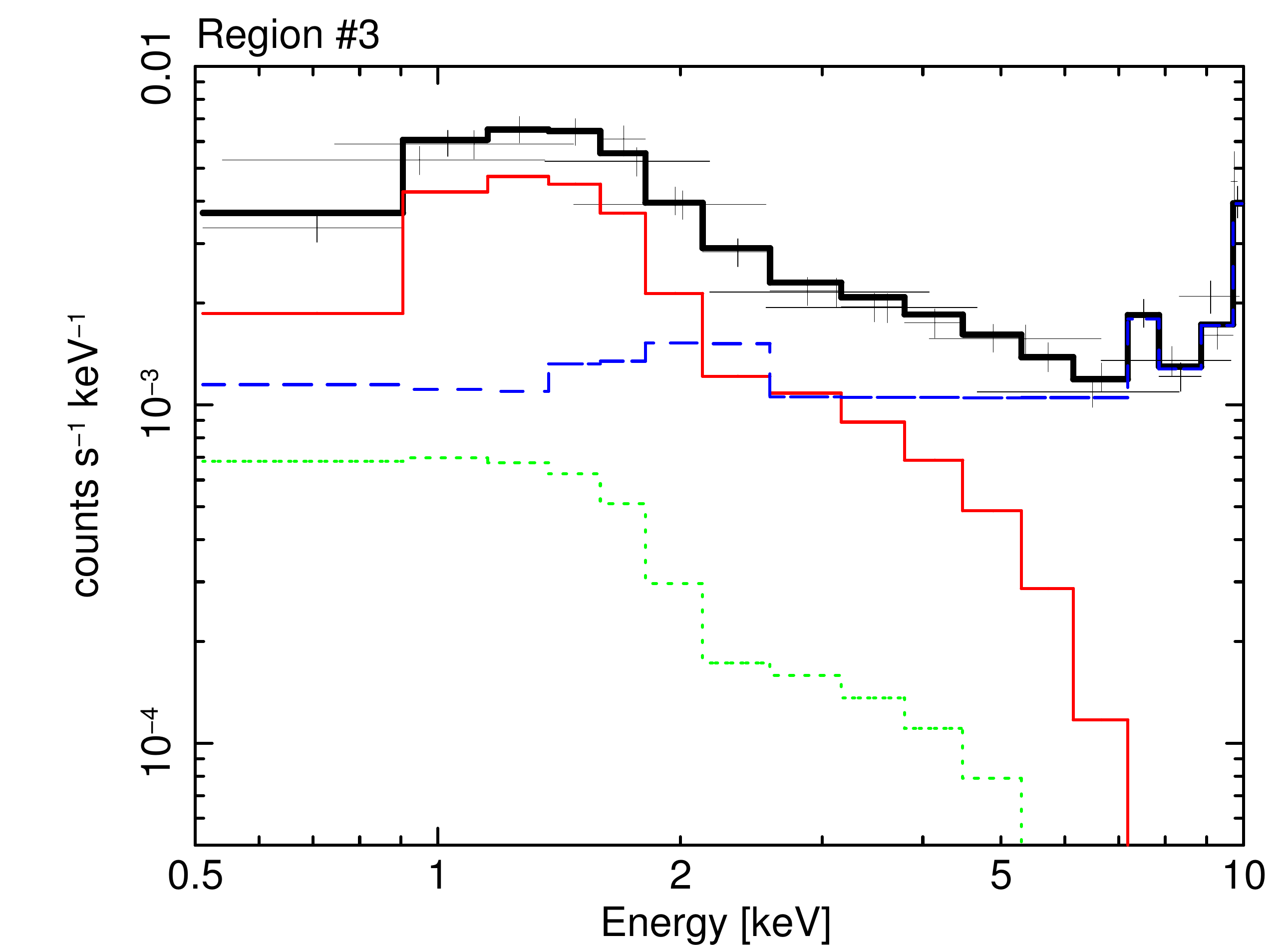}
 \includegraphics[width=.3\hsize,angle=-90]{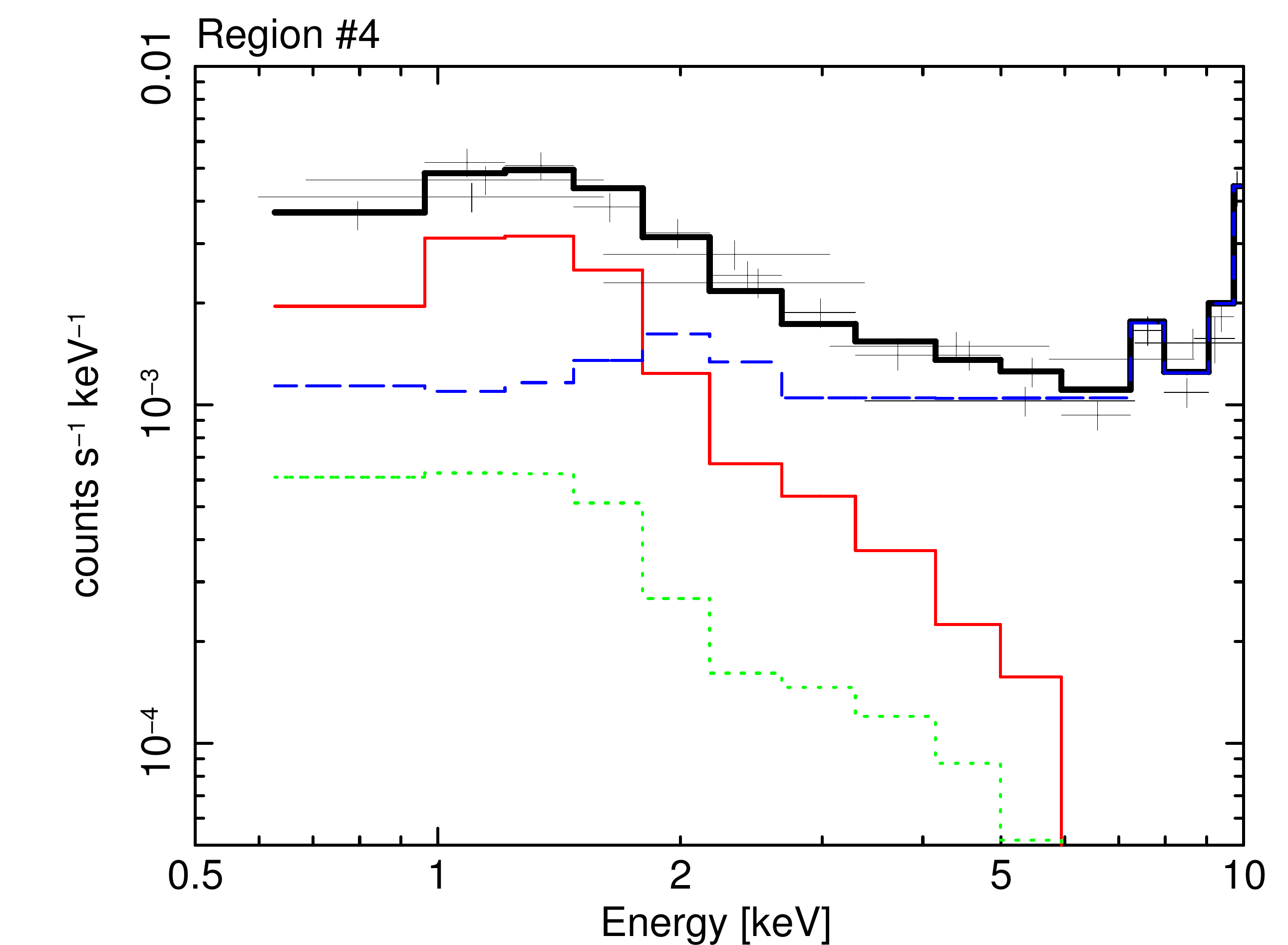}
 \includegraphics[width=.3\hsize,angle=-90]{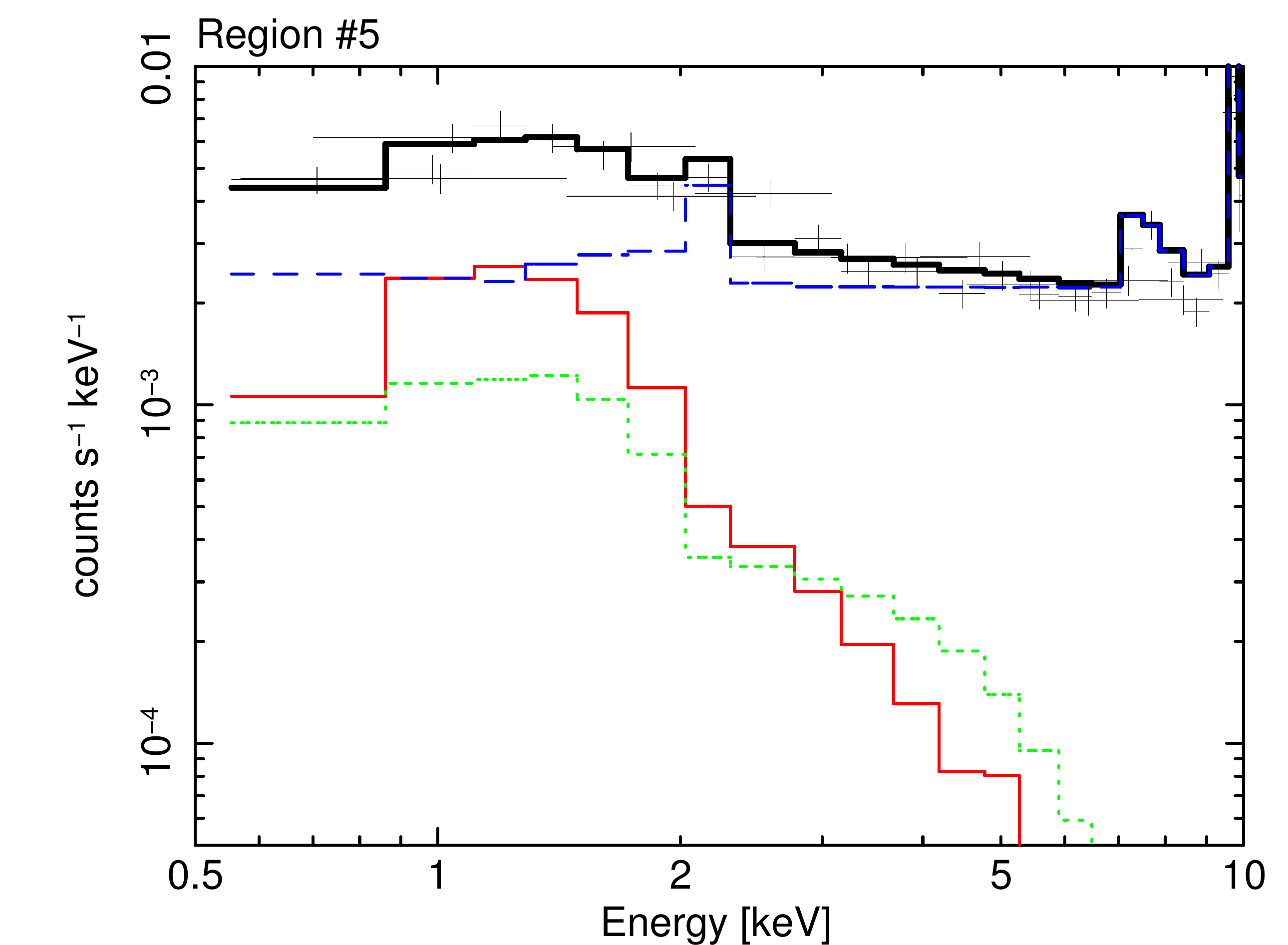}
 \caption{Spectral fitting results for the five regions shown in Fig.~\ref{fig:shock} (left panel). Data points are shown in black together with the best-fitting model. Different colors denote the components of the spectral model; \ie, the cluster emission (in solid red), the astrophysical background (in dotted green) and the instrumental background (in dashed blue). Whilst the three \obsid\ spectra were simultaneously fitted, the models for only one observation were reported in order to avoid confusion in the plot. The \cstatdof\ of the fits from region 1 to 5 are: 232/184, 135/127, 114/132, 115/113 and 224/182.}
 \label{fig:spectra}
\end{figure*}

The best-fitting spectra of the five regions shown in Fig.~\ref{fig:shock} are reported in Fig.~\ref{fig:spectra}. The spectral model components are depicted with different colors in the plots. To asses the impact of the systematic uncertainties of the background modeling to the estimates of the ICM temperature, we re-performed the spectral fits varying within $\pm1\sigma$ the normalization levels first of the instrumental background alone, and later of both the instrumental and astrophysical backgrounds. The results are summarized in Tab.~\ref{tab:regions15} and are consistent within $1\sigma$ with that reported in Fig.~\ref{fig:shock} (right panel). Finally, we mention that the drop of the \chandra\ effective area above 5 \kev\ makes the estimation of high temperatures critical with this instrument. In this respect, the errors on the high temperatures reported in Tab.~\ref{tab:regions15} may not reflect entirely the real range of  statistical and systematic uncertainties.

\begin{table}
 \centering
 \caption{Impact of the systematic uncertainties of the background modeling on the temperature estimates reported in Fig.~\ref{fig:shock} (right panel). Tests were performed varying within $\pm1\sigma$ the normalization level of the instrumental background (``NXB'') and of the astrophysical background (``sky''). Temperatures are reported in \kev\ units.}
 \label{tab:regions15}
  \begin{tabular}{lccccc} 
  \hline
  Region & Best fit & \multicolumn{2}{c}{NXB} & \multicolumn{2}{c}{NXB + Sky} \\
  & & $+1\sigma$ & $-1\sigma$ & $+1\sigma$ & $-1\sigma$ \\
  \hline
  1 & $4.4^{+2.8}_{-1.4}$ & $3.7^{+2.1}_{-1.0}$ & $5.4^{+4.1}_{-1.9}$ & $3.5^{+4.1}_{-1.5}$ & $5.1^{+3.4}_{-1.1}$ \\
  2 & $15.4^{+9.3}_{-4.8}$ & $14.1^{+7.7}_{-3.8}$ & $17.1^{+12.5}_{-5.1}$ & $15.4^{+9.8}_{-5.1}$ & $16.7^{+10.0}_{-5.1}$ \\
  3 & $12.2^{+3.9}_{-3.6}$ & $10.2^{+4.5}_{-2.3}$ & $13.6^{+3.5}_{-3.9}$ & $11.8^{+4.1}_{-3.4}$ & $13.2^{+3.5}_{-3.7}$ \\
  4 & $5.0^{+1.4}_{-0.9}$ & $4.5^{+1.2}_{-0.7}$ & $5.4^{+1.5}_{-1.1}$ & $4.8^{+1.4}_{-0.9}$ & $5.3^{+1.5}_{-1.0}$ \\
  5 & $2.5^{+0.7}_{-0.5}$ & $2.2^{+0.6}_{-0.3}$ & $2.7^{+0.9}_{-0.5}$ & $2.3^{+0.7}_{-0.4}$ & $2.8^{+0.8}_{-0.6}$ \\
  \hline
 \end{tabular}
\end{table}

\bsp	
\label{lastpage}
\end{document}